# Optomechanical transducers for gigahertz frequency ultrasound imaging

Lisa Hackett,[1] Chang Ge,[2] Alex Miera,[3] Brandon Smith,[1] and Matt Eichenfield[1,2,3]

1. Microsystems Engineering, Science and Applications, Sandia National Laboratories, Albuquerque, NM, USA
2. James C. Wyant College of Optical Sciences, University of Arizona, Tucson, AZ, USA
3. Electrical, Computer and Energy Engineering, University of Colorado, Boulder, CO, USA

Correspondence: Lisa Hackett (lahacke@sandia.gov) and Matt Eichenfield (matt.eichenfield@colorado.edu)

## Abstract

Ultrasound is a noninvasive, real-time, and therefore widely used imaging modality; yet its application in cellular and sub-cellular biology is significantly limited by rapidly increasing acoustic losses in aqueous solutions with decreasing wavelength. Here we introduce a nano-optomechanical cavity transducer platform to generate and detect ultrasound in aqueous solutions with a sub-micron acoustic wavelength. We analyze the full signal pathway through a combination of finite element method modeling and the coupled differential equations that describe the dispersive optomechanical interaction. Our findings project a signal-to-noise ratio in the thousands at ~5 GHz, limited by diffraction losses and thermal-acoustic noise. This work establishes a viable path towards optomechanical ultrasound systems capable of label-free imaging at cellular and sub-cellular length scales while also providing a broader framework for optomechanical crystal device operation in aqueous environments relevant to biochemical sensing, medical diagnostics, underwater acoustic sensing, and nanoscale imaging.

# Introduction

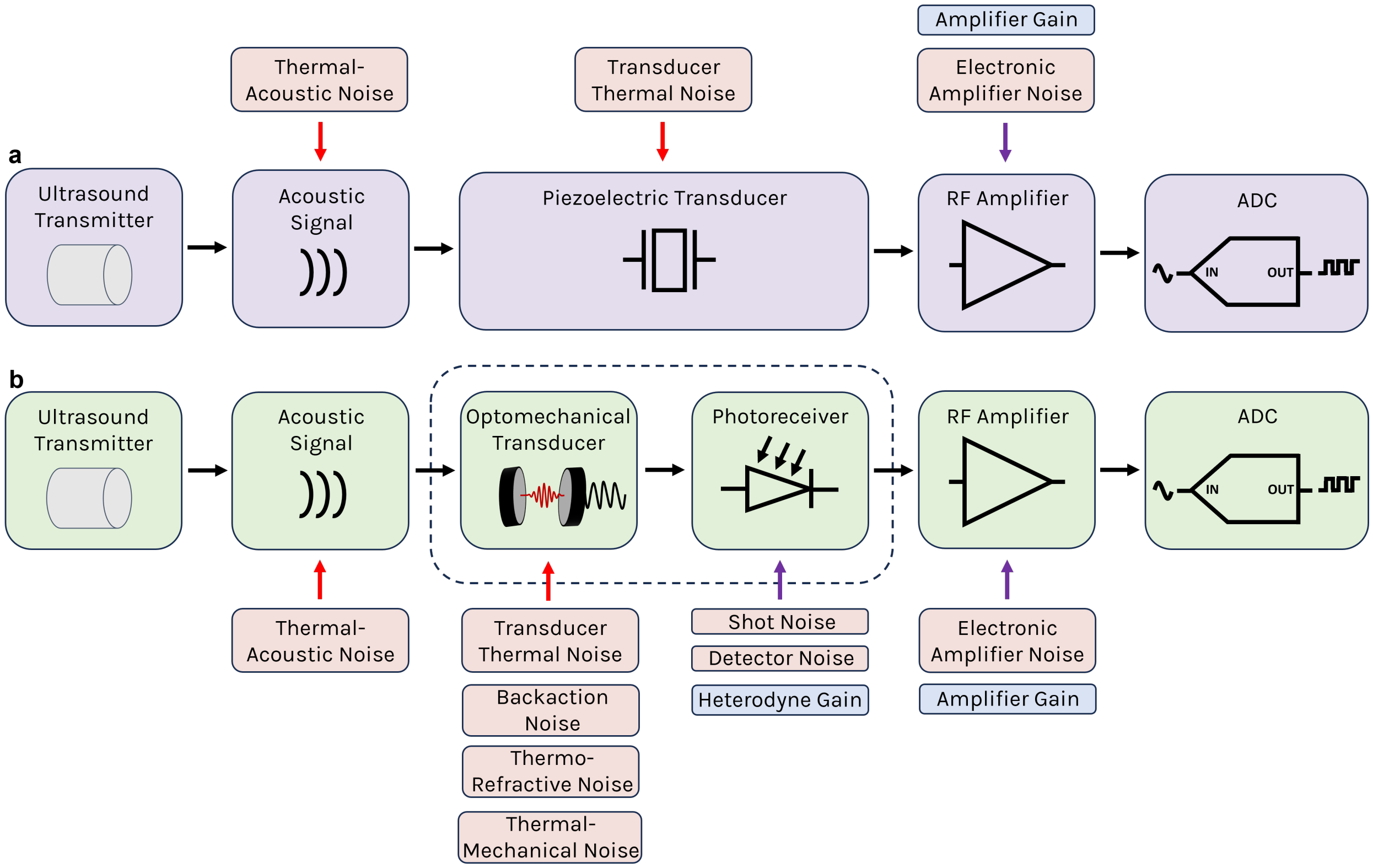

**Fig. 1: Ultrasound signal pathway.** Block diagrams of the ultrasound signal path for detection with (a) a piezoelectric transducer or (b) an optomechanical transducer. While both approaches must contend with acoustic-to-electrical conversion losses, optomechanical ultrasound detection occurs through a two-step transduction process (acoustic to optical followed by optical to electrical) where the radio frequency (RF) spectrum is obtained by optical heterodyne detection on a photodiode with the potential for large heterodyne gain.

Ultrasound could be a powerful method for label-free and real-time imaging of biological cells and sub-cellular structures; however, the application of ultrasound to these size scales is limited by fluidic viscous damping losses, which increase quadratically with the operating frequency. Conventionally, bulk piezoelectric transducers transmit and detect ultrasound signals via radio frequency (RF) transduction of acoustic waves. Scanning acoustic microscopy[1,2] has achieved gigahertz operating frequencies through techniques such as high temperature (60°C) operation where acoustic attenuation in water is decreased, or utilizing liquid helium at cryogenic temperatures, neither of which are universally compatible with live-cell in-vitro imaging[2,3].

A schematic of the canonical ultrasound signal pathway is shown in Fig. 1(a). The input-referred minimum detectable signal is characterized by a noise equivalent pressure (NEP), which defines the smallest pressure level that produces a signal equal to the noise level of the system on the output. For ultrasound systems operating in liquid water, the NEP is fundamentally limited by the pressure noise floor due to random thermal vibrations of water molecules, often referred to as the thermal-acoustic noise of the medium. To date, for piezoelectric ultrasound transducers, the acoustic-to-electrical transduction efficiency is substantially less than unity at frequencies exceeding 50 MHz, resulting in the minimum detectable signal being limited by the electronic noise floor of the requisite RF amplifier that follows the transducer.[4-7]

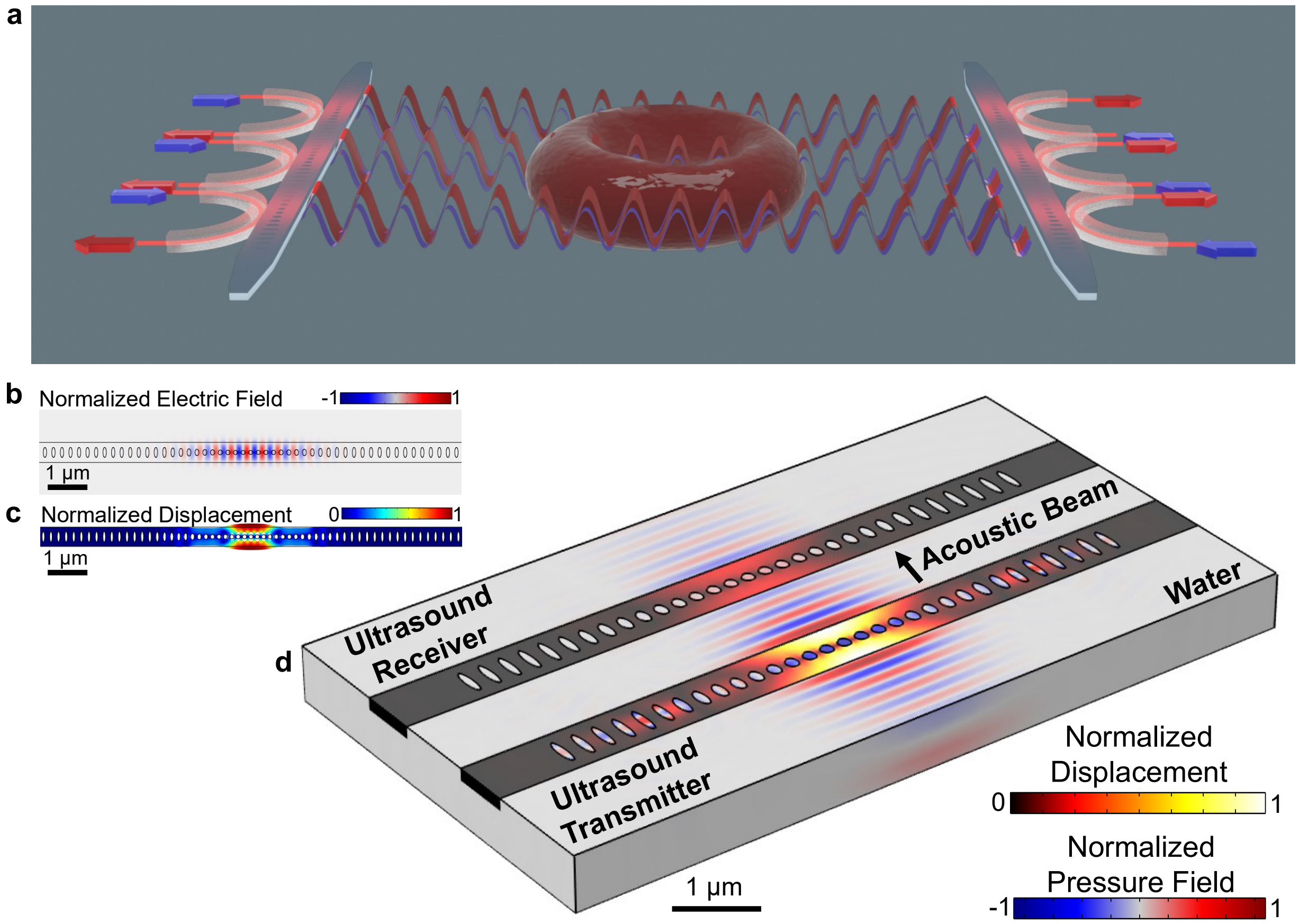

**Fig. 2: Optomechanical ultrasound transducers.** (a) An illustration of the proposed ~5 GHz ultrasound system in which OMC devices, submerged in an aqueous solution, are used to both generate and detect gigahertz acoustic signals. A linear array of optomechanical cavities is shown, which transmit and receive acoustic beams that pass through a red blood cell. (b) The modeled normalized electric field and (c) normalized displacement field for an OMC in air. (d) The result of a FEM model of the ultrasound transduction process for the OMC transmitter and receiver immersed in a viscous fluid.

While piezoelectric[8-11] and capacitive[12,13] machined ultrasound transducers have been proposed for acoustic microscopy with an improved minimum detectable signal, they are typically limited to frequencies <100 MHz. Super-resolution ultrasound imaging[14] overcomes the resolution limitations of traditional ultrasound, but relies on the injection or genetic engineering of contrast agents.[15,16] Optical ultrasound detectors such as Fabry-Perot cavities, π-phase-shifted Bragg gratings, and whispering gallery mode microcavities can provide improved sensitivity at acoustic frequencies >100 MHz[17-19] due to high quality factor optical resonance and the two-step transduction process (acoustic to optical followed by optical to electrical), which offers more opportunities for noise reduction and design optimization. Brillouin microscopy provides label-free mechanical imaging by focusing light into a sample and detecting Brillouin-scattered photons with a spatial resolution that is primarily determined by optical diffraction.[20,21] Picosecond ultrasonics has demonstrated single-cell imaging with a ~2 μm lateral resolution using ultrafast pump-probe optics, but the cells had to be cultured directly on an absorbing transducer.[22] Photoacoustic microscopy has achieved sub-micron resolution, but operates based on optical absorption in the sample.[23] In contrast, our approach uses nanoscale optomechanical transducers to directly generate and detect propagating gigahertz frequency ultrasound signals in water, such that the optical transduction and sample interrogation are inherently separated. This provides a path towards high resolution ultrasound imaging based on direct acoustic contrast using chip-scale architectures and arrays.

An optomechanical detection system consists of coupled mechanical and optical modes in a cavity. The pressure field in water drives the mechanical mode and its displacement modulates the optical resonance frequency and generates sidebands. The power exiting the cavity, which contains the transmitted optical sidebands and the carrier, is then incident on a photodetector, where the RF spectrum is obtained by optical heterodyne detection on a photodiode. The optical power oscillates at the RF signal frequency with an amplitude that is proportional to the square root of the product of the transmitted optical carrier power and sideband power, such that the carrier acts as a local oscillator to enable heterodyne gain. Optomechanical detection then provides a two-stage acoustic-to-optical-to-electrical transduction process where the optical-to-electrical heterodyne gain can overcome the limitation of a finite transduction efficiency. Figure 1(b) summarizes the potential limiting noise sources of optomechanical ultrasound detection. To the best of our knowledge, a similarly comprehensive noise analysis has not been reported for piezoelectric detection and is less relevant at the frequencies considered here where the NEP is known to be limited by electronic amplifier noise.

In this article, we design an optomechanical transduction platform for gigahertz frequency ultrasound imaging with optomechanical crystal (OMC) transmitters and receivers operating in aqueous solutions. OMCs are state-of-the-art displacement sensors[24,25] with exquisite sensitivity that is achieved by co-localization of optical and mechanical resonances within a sub-cubic-micron volume. The same OMC structure can also be optically pumped into large amplitude oscillation, causing it to operate as an ultrasound transmitter. We design the system and determine the achievable signal-to-noise ratio (SNR) using the coupled differential equations that describe the dispersive optomechanical interaction along with finite element method (FEM) modeling. Overall, in this work, we show that despite significant acoustic losses in aqueous solutions, the SNR of optomechanical ultrasound transducers operating near 5 GHz can be exceptionally large. These results suggest that our approach is a viable path forward to enabling high-resolution ultrasound imaging with potential applications in fundamental biological studies of cellular mechanics, detection of viral load, disease diagnosis, and much more.[26-28]

# Results

### Layout and operating principle for optomechanical ultrasound

In our proposed system, illustrated in Fig. 2(a), a single OMC transmitter and receiver pair, or a linear array, could operate in a setup similar to conventional transmission ultrasound.[29-31] 2D or 3D imaging could be achieved by translating the chip or sample, including through optical or acoustic trapping methods.[32,33] The proposed system targets narrowband transmission imaging rather than broadband pulse-echo or photoacoustic imaging. In this modality, resolution is determined primarily by the acoustic wavelength and beam profile rather than the pulse bandwidth. Future designs could broaden the mechanical bandwidth by increasing acoustic radiation loading into the fluid. The modeled electric and displacement fields for such OMC devices in air are shown in Fig. 2(b) and Fig. 2(c), respectively, while Fig. 2(d) shows the modeled transduction process in an aqueous solution. The mechanical displacement of the ultrasound transmitter generates a propagating acoustic beam that drives displacement in the receiver, ultimately modulating the detected optical signal.

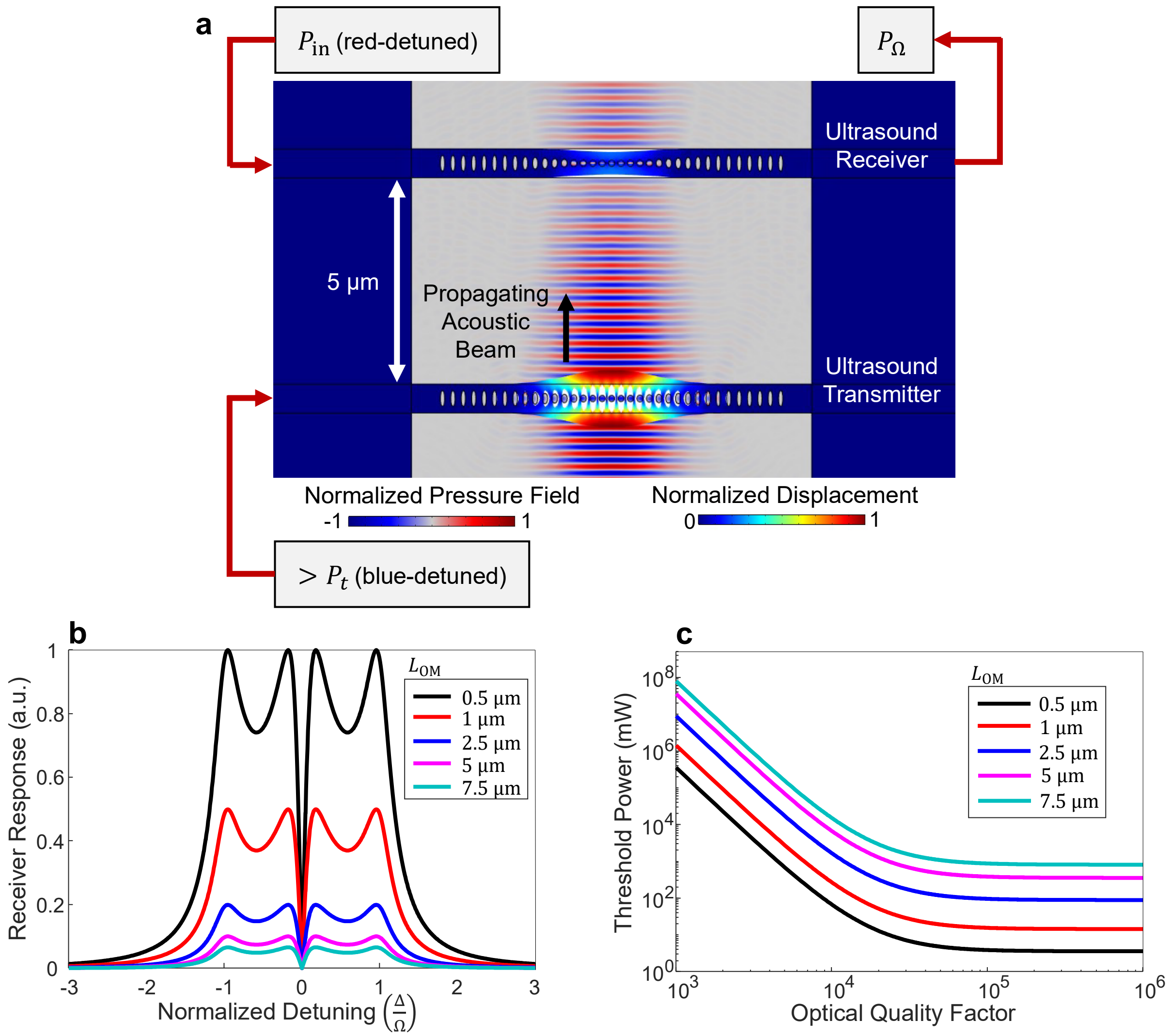

**Fig. 3: Overview of the optomechanical ultrasound system.** (a) In the proposed optomechanical ultrasound system, two aligned silicon OMCs with a fixed gap, $g$, between them are submerged in an aqueous solution. The acoustic beam, generated by optical pumping with an input power above a threshold value $P_t$, propagates and induces oscillations in an OMC receiver that modulates an optical carrier signal with input optical power $P_{in}$. (b) The modeled receiver response as a function of detuning, $\Delta$, and the effective optomechanical coupling length, $L_{OM}$. As can be seen, enhancing the receiver response requires operating at the optimal detuning point and minimizing $L_{OM}$. (c) The modeled required threshold optical power to reach optomechanical regenerative oscillations in the transmitter as a function of the optical quality factor and $L_{OM}$.

Additional details are shown in Fig. 3(a). In each OMC, an optical mode described by an electric field amplitude $A$ and a mechanical breathing mode described by a displacement amplitude $a$ mutually couple. To achieve a large acoustic amplitude in the water, the transmitter is optically pumped above a threshold power, $P_t$, into the regime of optomechanical regenerative oscillations. [34-38] For a typical silicon OMC mechanical frequency of $\Omega_m/2\pi$ = 5 GHz, the generated acoustic beam in the water has a wavelength, $\Lambda$, of approximately 300 nm. The ultrasound receiver operates based on the principle of dispersive optomechanical coupling where the resonance frequency of the optical mode, $\omega_0$, is a function of the mechanical displacement according to $\omega_0(a) = \omega_0 - a\frac{\omega_0}{L_{\mathrm{OM}}}$ where $L_{\mathrm{OM}}$ is the effective optomechanical length given by $L_{\mathrm{OM}}^{-1} = \frac{1}{\omega_0}\frac{d\omega_0}{da}$. The receiver then operates based on the change in $a$ due to the impinging acoustic beam.

To first order, the coupled differential equations that describe the dispersive optomechanical interaction in each OMC are[25]

$$\dot{A}(t) = \left(-\frac{\gamma}{2} - i\omega_0\left(1 - \frac{a(t)}{L_{\mathrm{OM}}}\right)\right)A(t) + \sqrt{\gamma_w}se^{-i\omega t} \tag{1}$$

and

$$\ddot{a}(t) + \Gamma\dot{a}(t) + \Omega_m^2 a(t) = \frac{F(t)}{m_{\mathrm{eff}}}. \tag{2}$$

Equation (1) describes the time evolution of $A$ driven by an input field with a power given by $P_{\mathrm{in}} = |s|^2$ and a frequency $\omega$ where $\gamma = \gamma_{\mathrm{int}} + 2\gamma_w$ is the total optical loss rate, $\gamma_{\mathrm{int}}$ is the intrinsic cavity loss rate, and $\gamma_w$ is the waveguide-cavity coupling rate. Equation (2) describes the dynamics of the mechanical resonator's displacement driven by a force $F(t)$. The mechanical resonator has an effective mass, $m_{\mathrm{eff}}$, and a total damping rate $\Gamma = \Gamma_{\mathrm{int}} + \Gamma_f$, where $\Gamma_{\mathrm{int}}$ is the intrinsic decay rate of the resonator and $\Gamma_f$ is the fluidic damping rate. As described in detail in Supplementary Note 1, from Eqns. (1-2) we can determine expressions for the receiver's output power oscillating at the RF frequency, $P_\Omega$, and the necessary optical input power to the transmitter to reach the threshold for optomechanical regenerative oscillations, $P_t$.

Optomechanical coupling generates a set of sidebands in the optical cavity that can be observed by measuring the optically transduced RF spectrum through optical heterodyne detection on a photodiode. Assuming that the optomechanical modulation, quantified by a dimensionless index $\beta = a_0\omega_0/L_{\mathrm{OM}}\Omega$ is weak, only the first-order sidebands contribute and the transmitted signal oscillating at $\Omega$, $T_\Omega$, is

$$T_{\Omega} = A_c \cos(\Omega t) + A_s \sin(\Omega t) \quad (3)$$

$$= \cos(\Omega t)\, \beta \gamma_w^2 \left(\frac{2\gamma}{\gamma^2 + 4\Delta^2}\right)\left(\frac{\Omega}{\left(\frac{\gamma}{2}\right)^2 + (\Omega - \Delta)^2} - \frac{\Omega}{\left(\frac{\gamma}{2}\right)^2 + (\Omega + \Delta)^2}\right)$$

$$+ \sin(\Omega t)\, \beta \gamma_w^2 \left(\frac{4\Omega}{\gamma^2 + 4\Delta^2}\right)\left(\frac{\Omega - \Delta}{\left(\frac{\gamma}{2}\right)^2 + (\Omega - \Delta)^2} - \frac{\Omega + \Delta}{\left(\frac{\gamma}{2}\right)^2 + (\Omega + \Delta)^2}\right)$$

where $A_c$ is the in-quadrature component, $A_s$ is the in-phase component, and $\Delta = \omega - \omega_0$ is the detuning between the optical carrier and optical cavity resonance frequencies. $P_{\Omega}$ can be determined from Eqn. (3) according to $P_{\Omega} = P_{\text{in}}\sqrt{A_c^2 + A_s^2}$. Figure 3(b) shows a plot of the normalized OMC receiver response as a function of $\Delta$ for different $L_{\text{OM}}$ values. As can be seen, enhancing the receiver response requires operating at the optimal detuning point and minimizing $L_{\text{OM}}$.

The threshold optical input power, $P_t$, to achieve optomechanical regenerative oscillations for the OMC ultrasound transmitter is given by

$$P_t = \frac{\Gamma m_{\text{eff}} \Omega_m L_{\text{OM}}^2}{\omega_0}\left[\left(\frac{2\gamma_w}{\gamma^2 + 4\Delta^2}\right)\left(\frac{\gamma}{\left(\frac{\gamma}{2}\right)^2 + (\Omega_m - \Delta)^2} - \frac{\gamma}{\left(\frac{\gamma}{2}\right)^2 + (\Omega_m + \Delta)^2}\right)\right]^{-1}. \quad (4)$$

While $P_t$ decreases with increasing mechanical quality factor, $Q_m$, in an optomechanical ultrasound transmitter the objective is to radiate acoustic energy into the water to generate the acoustic beam, which drastically restricts the maximum value for $Q_m$. Figure 3(c) shows a plot of $P_t$ as a function of $Q_{\text{opt}}$ for different $L_{\text{OM}}$ on a log-log plot for the case where $Q_m = 10$. Despite low $Q_m$, an experimentally achievable $P_t$ of 15 mW or less can be generated if $L_{\text{OM}}$ is 1 µm and $Q_{\text{opt}}$ is at least $1.4 \times 10^5$.

**Optomechanical design**

To realize high frequency optomechanical ultrasound transducers in aqueous solutions, a large optical quality factor, $Q_{\text{opt}}$, must be achieved. To mitigate optical absorption in water, we design the silicon OMCs to operate at 1300 nm where the absorption coefficient for water is approximately 1.1 $\text{cm}^{-1}$,[39] which is significantly less than the value of approximately 10 $\text{cm}^{-1}$ at 1550 nm, which is the conventional wavelength for OMC design.[25,40] We also increase confinement of the optical cavity mode in the silicon to reduce evanescent overlap with the lossy surrounding environment. One configuration to deliver light to the optical cavity, which we utilize for our analysis here, is through edge-coupled waveguides (see Supplementary Note 2).

The OMC design follows previous studies[25,40-42] and is modeled in water with a refractive index of 1.316 and an extinction coefficient of $1.4\text{x}10^{-5}$.[43] The mirror unit cell is shown in Fig. 4(a) with the computed bandstructure in water shown in Fig. 4(b). The defect unit cell and bandstructure are shown in Fig. 4(c) and Fig. 4(d), respectively. The optical cavity mode (Fig. 4(e)), has a modeled intrinsic optical quality factor, $Q_{\text{int}} = \omega_0/\gamma_{\text{int}}$, of $1.2 \times 10^7$ in air, $6.6 \times 10^6$ in water with no extinction coefficient, and $8.3 \times 10^5$ in water with an extinction coefficient. While absorption in water significantly reduces $Q_{\text{int}}$, experimentally it is already limited to $\sim 10^6$ in air due to surface scattering.[40] The mechanical modes and phononic bandstructures for the mirror and defect unit cells in air are shown in Fig. 4(f) and Fig. 4(g), respectively. We utilize FEM modeling to confirm that the mechanical stop band is preserved when the OMC is immersed in water, as described in Supplementary Note 3.

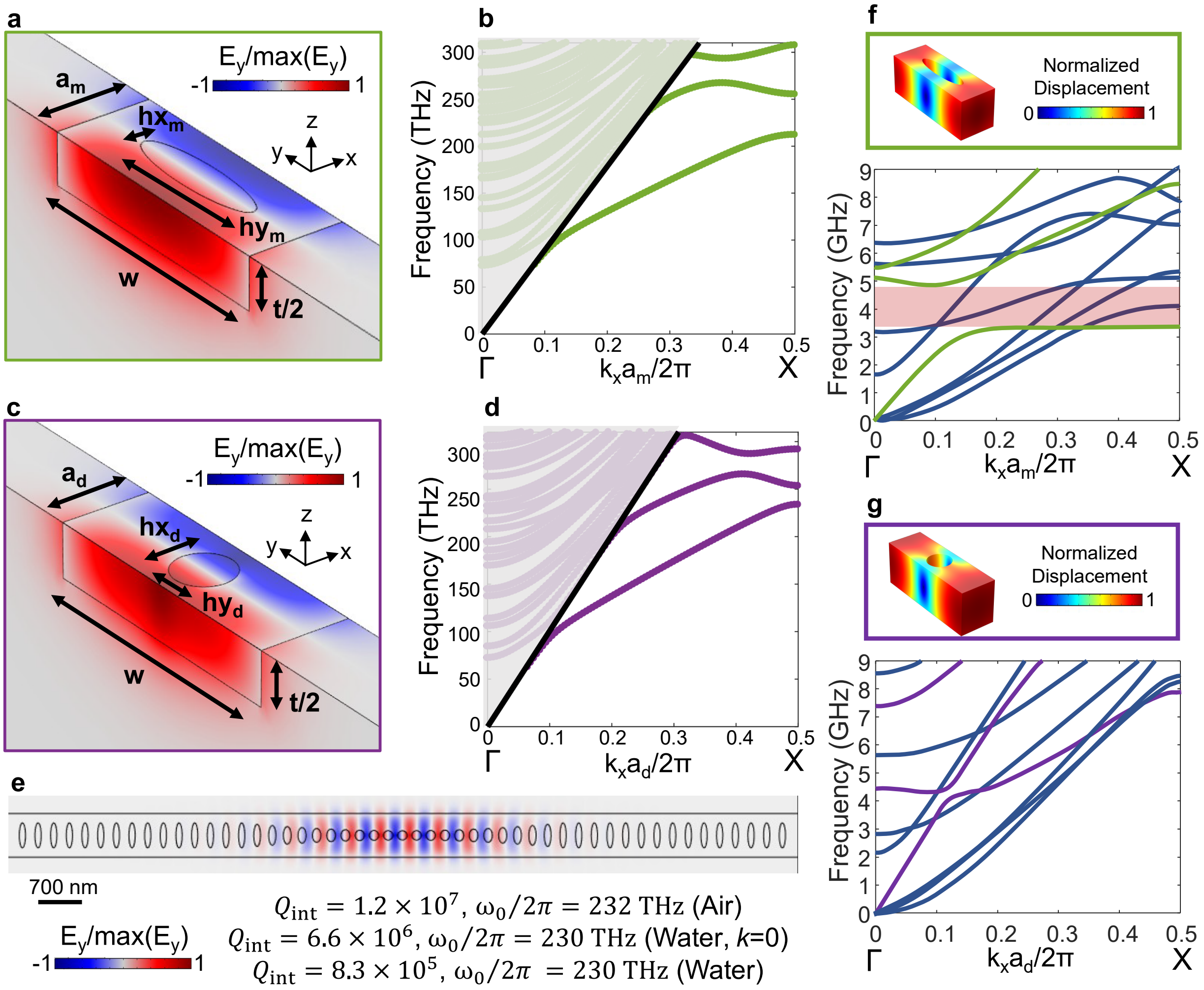

**Fig. 4: Optical and mechanical design of the submerged OMC.** (a) The mirror unit cell of the OMC with (b) the corresponding photonic bandstructure. (c) The defect unit cell with (d) the corresponding bandstructure. (e) The modeled optical mode in the OMC with the simulated $Q_{int}$ values. Normalized displacements and phononic bandstructures in air for the (f) mirror and (g) defect unit cells.

## Acoustic beam generation

Here we characterize the mechanical oscillator of the OMC and assess acoustic beam generation using a FEM acoustic-structure interaction model that treats the fluid as a viscous medium. To excite the breathing mode, a driving force with a Gaussian distribution is applied laterally along the length of the beam to follow the known displacement (Fig. 2(c)). The displacement and pressure fields from a 3D FEM study are shown for two different perspectives in Fig. 5(a) and 5(b). From this model, the optomechanical coupling length is 1 μm and the effective mass is 307 fg (Supplementary Note 4), demonstrating that low values can be obtained even in the viscous fluid.

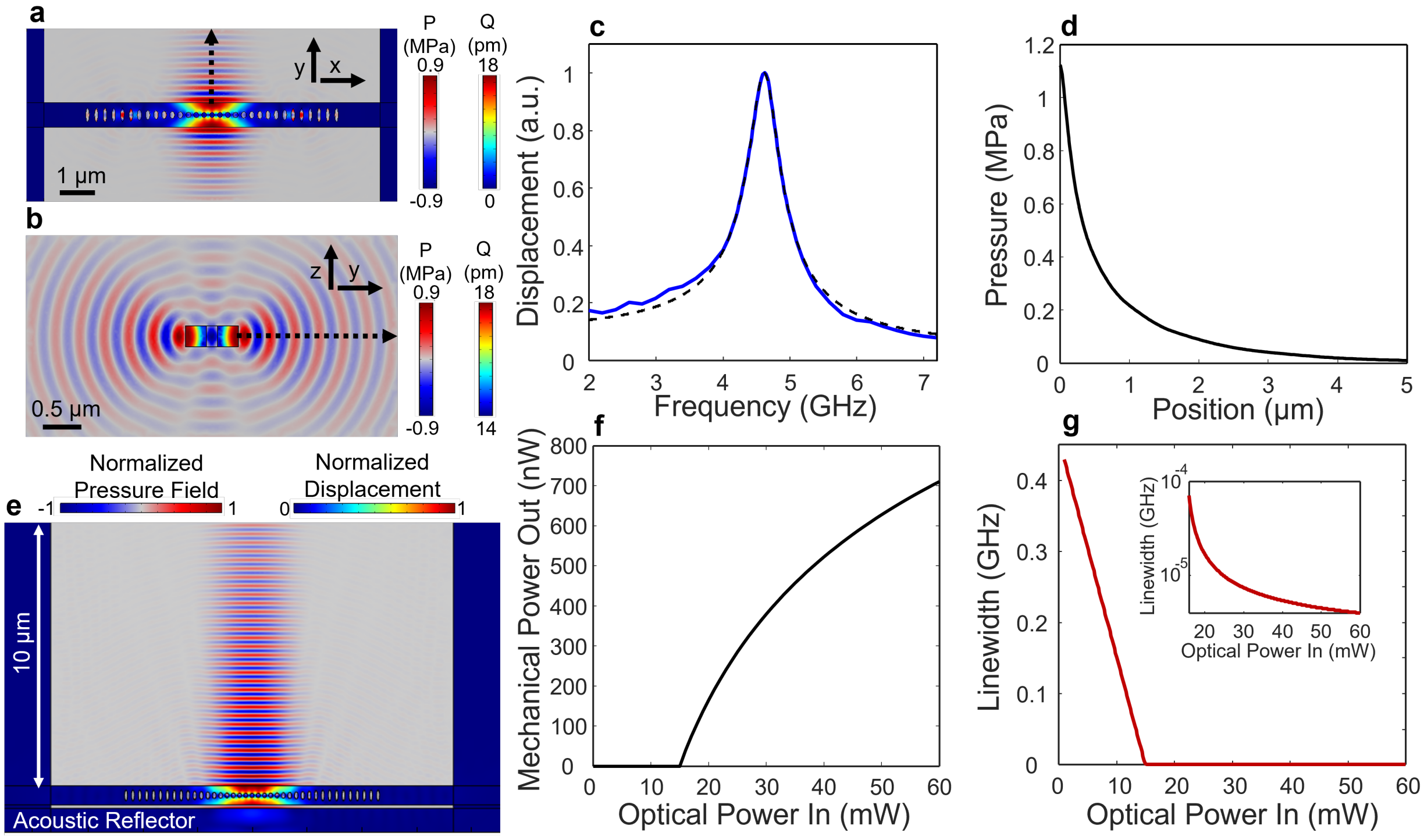

**Fig. 5: Acoustic beam properties.** The properties of the mechanical oscillator and generated acoustic beam are determined through a 3D FEM model of the displacement and pressure field as shown in (a) and (b). (c) The $Q_m$ from the 3D FEM model. (d) The absolute total acoustic pressure as a function of position away from the OMC transmitter, as indicated by the dashed black lines in (a) and (b). (e) A modeled acoustic beam from a 2D FEM model with an acoustic reflector. The modeled OMC transmitter (f) mechanical output power due to self-sustained optomechanical regenerative oscillations and (g) the corresponding mechanical linewidth, $\Gamma_{eff}/2\pi$. For these calculations, we set $P_t = 15$ mW, $Q_{opt} = 1.5 \times 10^5$, $\omega_0/2\pi = 231$ THz, $\Omega_m/2\pi = 4.6$ GHz, $Q_m = 10$, $m_{eff} = 300$ fg, and $L_{OM} = 1$ µm. The inset plots the mechanical linewidth on a log scale above the threshold optical input power, $P_t$.

The generated acoustic beam experiences Stokes and diffraction losses. The Stokes-Kirchhoff attenuation coefficient, $\alpha$, is $\alpha = \frac{\Omega^2}{2\rho c^3}\left(\frac{4}{3}\eta_S + \zeta\right)$ where $\eta_S$ is the shear viscosity, $\rho$ is the density of the fluid, and $\zeta$ is the bulk viscosity.[44] At an operating frequency of $\Omega/2\pi = 4.6$ GHz, $\alpha \sim 4$ dB/µm. While the transmitter generates a Gaussian acoustic beam with a Rayleigh range of approximately 65 µm, diffraction losses along the thickness of the beam, captured in Fig. 5(a) and Fig. 5(b), are significant. The simulated $Q_m$ for the mechanical oscillator in the viscous fluid is $Q_m = 10$ (Fig. 5(c)) and does not change with reduced dynamic viscosity, which indicates that the viscous dissipation in the fluid is negligible compared with acoustic radiation losses. Assuming $Q_m = 3000$ in air,[45] this implies that >99% of the acoustic power loss is radiated into the fluid.

Figure 5(d) plots the absolute total acoustic pressure as a function of position along the black dotted cut lines shown in Fig. 5(a) and Fig. 5(b). As can be seen, the pressure amplitude of the acoustic beam decreases by over an order of magnitude 5 µm away from the OMC transmitter, emphasizing the need to maximize transmitter output power and minimize the receiver noise floor. A unidirectional transmitter could increase the pressure amplitude at the receiver for a fixed optical input power. As shown in Fig. 5(e), adjusting the spacing between the OMC transmitter and the silicon produces an acoustic reflector with constructive interference towards the receiver.

For a blue detuned pump above an optical threshold input power, $P_t$, (Eqn. (4)), a large amplitude acoustic beam can be achieved through self-sustained optomechanical oscillations. Supplementary Note 5 describes our calculation for the transmitter's mechanical output power, $P_{out}$, and the effective mechanical linewidth, $\Gamma_{eff}$, in the idealized linear cavity limit, which are plotted as a function of the optical input power, $P_{in}$, in Figures 5(f) and 5(g), respectively. An optical input power of $P_{in} = 50$ mW results in a $P_{out}$ of 630 nW and an $\Gamma_{eff}/2\pi$ of 4 kHz. Operation at these optical powers will require mitigation of nonlinear optical loss, particularly free-carrier absorption, using established carrier extraction or carrier lifetime engineering approaches in silicon photonics, including photonic crystal cavities.[46-49] The reduced $Q_m$ in water can be compensated with an increased $P_{in}$, but the resulting intracavity optical intensity can lead to unintended device dynamics such as self-pulsing.[50] In Supplementary Note 6, we show that immersion in water increases the thermal dissipation by more than an order of magnitude, which substantially raises the optical power threshold for self-pulsing to >80 mW, enabling suppression of this effect.

### Receiver performance

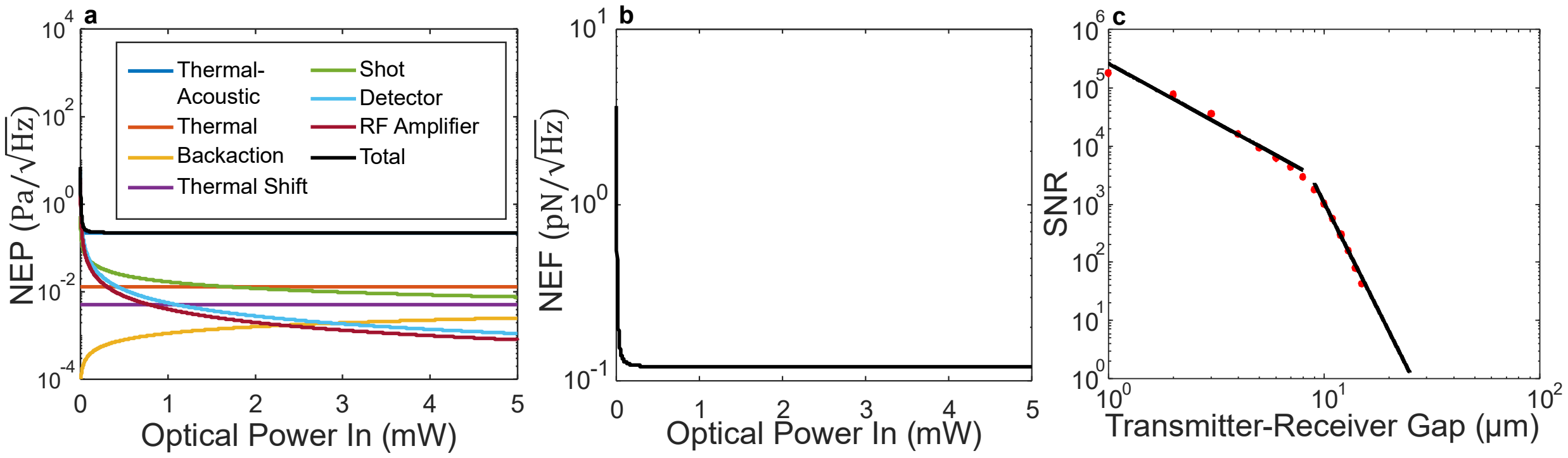

**Fig. 6: OMC receiver performance.** The calculated (a) noise equivalent pressure and (b) noise equivalent force as a function of optical input power to the OMC receiver for a mechanical operating frequency of 4.6 GHz, $m_{eff} = 300\ fg$, $Q_m = 10$, optical wavelength of 1300 nm, $Q_{opt} = 1.5 \times 10^5$, $L_{OM} = 1\ \mu m$, photodiode responsivity, $R$, of 0.95 $A/W$, photodiode noise equivalent power, $NEP_{det}$, of $1 \times 10^{-12}\ W/\sqrt{Hz}$, and a RF amplifier noise factor, $F_n$, of 2. (c) The calculated SNR as a function of the gap between the transmitter and receiver on a log-log scale with a broken power law fit. For gaps of 1 to 8 µm, the fitting gives an exponent of approximately -2 while for gaps from 9 to 15 µm, the fitting gives an exponent of approximately -7.

To characterize the OMC ultrasound receiver performance, we calculate its noise equivalent pressure, NEP, which determines the minimum detectable signal. The total NEP, expressed in units of Pa/√Hz, includes contributions from the thermal-acoustic noise from the medium, $p_a$, the intrinsic thermal noise of the mechanical oscillator, $\mathrm{p_{th}}$, backaction noise, $\mathrm{p_{BA}}$, thermo-refractive noise, $\mathrm{p_{TR}}$, thermal-mechanical noise, $\mathrm{p_{TM}}$, shot noise, $\mathrm{p_{SN}}$, detector noise, $\mathrm{p_{det}}$, and RF amplifier noise, $\mathrm{p_{amp}}$, according to

$$\mathrm{NEP} = \sqrt{\mathrm{p_a^2 + p_{th}^2 + p_{BA}^2 + p_{TR}^2 + p_{TM}^2 + p_{SN}^2 + p_{det}^2 + p_{amp}^2}}. \quad (5)$$

The NEP derivation and analysis is given in Supplementary Notes 7-9. Here we consider receiver performance independently of the source with transmitter noise discussed in Supplementary Note 7. In Fig. 6(a), the calculated NEP is plotted as a function of the optical input power to the receiver for a mechanical operating frequency of 4.6 GHz. The minimum NEP is 0.2 $\mathrm{Pa/\sqrt{Hz}}$ and is limited by thermal-acoustic noise. For a measurement bandwidth of 100 Hz, the minimum detectable signal is 2 Pa.

For comparison, a NEP of 1.3 $\mathrm{mPa/\sqrt{Hz}}$, limited by thermal-acoustic noise, was experimentally demonstrated for an optomechanical ultrasound detector at 3-30 MHz.[51] A scaling analysis, reported in Supplementary Note 10, predicts that the thermal-acoustic NEP, $p_a$, increases linearly with increasing mechanical frequency, which is consistent with our ~150X higher operating frequency (4.6 GHz vs 30 MHz) and NEP (0.2 $\mathrm{Pa/\sqrt{Hz}}$ vs 1.3 $\mathrm{mPa/\sqrt{Hz}}$). Supplementary Note 10 also describes ways to reduce $p_{\mathrm{a}}$ itself. To compare to detectors of different geometries, we use the noise equivalent force (NEF), which is the NEP multiplied by the effective detector area, σA. The minimum NEF is 0.1 $\mathrm{pN/\sqrt{Hz}}$ (Fig. 6(b)), which shows how this approach can lead to sufficient SNR despite the ultracompact detector area of 270 nm x 10 μm. Overall, these results suggest that gigahertz frequency ultrasound detectors are best suited for high resolution miniaturized systems with short propagation lengths.

Ultimately, we want to know if the receiver can produce a signal that exceeds the minimum detectable signal. We utilize our full 3D transduction model to determine the receiver's mechanical displacement, $a_{Rx}$, which is converted to pressure using the mechanical response function (Supplementary Note 7). The SNR is obtained by dividing by the minimum detectable signal. A plot of the SNR as a function of the transmitter-receiver gap, along with a broken power law fitting, is shown in Fig. 6(c) for 50 mW of optical input power to the transmitter. The SNR is >170,000 at a gap of 1 μm, >9,000 at a gap of 5 μm, and >1,000 at a gap of 10 μm. The SNR can then be extremely large with heterodyne gain enabling detection at the thermal-acoustic noise floor at 4.6 GHz. Importantly, the large calculated SNR does not require the full idealized OMC transmitter output. Because acoustic pressure and SNR scale approximately as $\sqrt{P_{\mathrm{mech}}}$, even a 10,000X reduction in mechanical output power would achieve a SNR exceeding 10 at a 10 μm gap. Moreover, the optomechanical receiver does not require an OMC source and could instead be paired with a piezoelectric transducer driven by an RF microwave source. Reducing the effective optomechanical length or engineering reduced diffraction losses, such as by increasing the OMC thickness, could improve the SNR at larger gaps. Supplementary Notes 11 and 12 discuss preliminary object imaging through beam occlusion and an alternative transduction approach where OMC devices in air generate and detect acoustic signals in an aqueous solution.

# Discussion

Here we have presented and evaluated optomechanical transducers for an ultrasound imaging platform based on the generation and detection of acoustic signals in water with a sub-micron wavelength. The entire ultrasound transduction process was modeled and the results suggest that our approach should allow high-resolution imaging of sub-cellular biological structures with linear imaging and could be extended to the scale of proteins with further work or the use of nonlinear acoustic imaging techniques.[52] Extensions of this work will examine optimal device configurations and operations for imaging, including experimental validation of our modeled transducer performance. Acoustic beam engineering to reduce diffraction losses would have a significant impact in improving this approach. The geometry of the silicon in the mechanical cavity region can potentially be engineered to form tailored beam profiles, such as focused beams with defined focal regions, thereby extending the approach to imaging modalities where a collimated beam is not optimal. The OMC design optimization for aqueous solutions has additional applications such as ultrasensitive mass detection of biological samples. The nano-optomechanical platform is also conducive for integration in microfluidic systems.

To the best of our knowledge, the heterodyne gain intrinsic to optomechanical detection has not been discussed in the literature and has implications beyond ultrasound sensing. While heterodyne gain comes with a noise factor of 2, homodyne detection or phase-sensitive parametric amplification could be incorporated into optomechanical readout to achieve a noise factor of 1.[53] Overall, our results establish nanoscale optomechanical transducers as a promising platform towards sub-micron resolution ultrasound imaging in aqueous environments, with large SNRs over micron-scale distances.

## Acknowledgements

This work was supported by the Laboratory Directed Research and Development program at Sandia National Laboratories, a multimission laboratory managed and operated by National Technology and Engineering Solutions of Sandia LLC, a wholly owned subsidiary of Honeywell International Inc. for the U.S. Department of Energy's National Nuclear Security Administration under contract DE-NA0003525. This work was performed, in part, at the Center for Integrated Nanotechnologies, an Office of Science User Facility operated for the U.S. Department of Energy (DOE) Office of Science. This paper describes objective technical results and analysis. Any subjective views or opinions that might be expressed in the paper do not necessarily represent the views of the U.S. Department of Energy or the United States Government.

## Author contributions

L.H. and M.E. performed the design, modeling, and analysis for the optomechanical ultrasound transducer submerged in water with input from C.G., A.M., and B.S. The dry cavity approach was designed, modeled, and analyzed by C.G., L.H., and M.E. with input from A.M. and B.S. All authors contributed to the discussion of the results and approved the final manuscript.

## Data availability

The datasets generated and analyzed during the current study are available from the corresponding author upon reasonable request.

## Conflict of interest

The authors declare that they have no conflict of interest.

# Supplementary Information
# Optomechanical transducers for gigahertz frequency ultrasound imaging
# L. Hackett et al.

## Supplementary Note 1. The dispersive optomechanical interaction in optomechanical ultrasound transducers

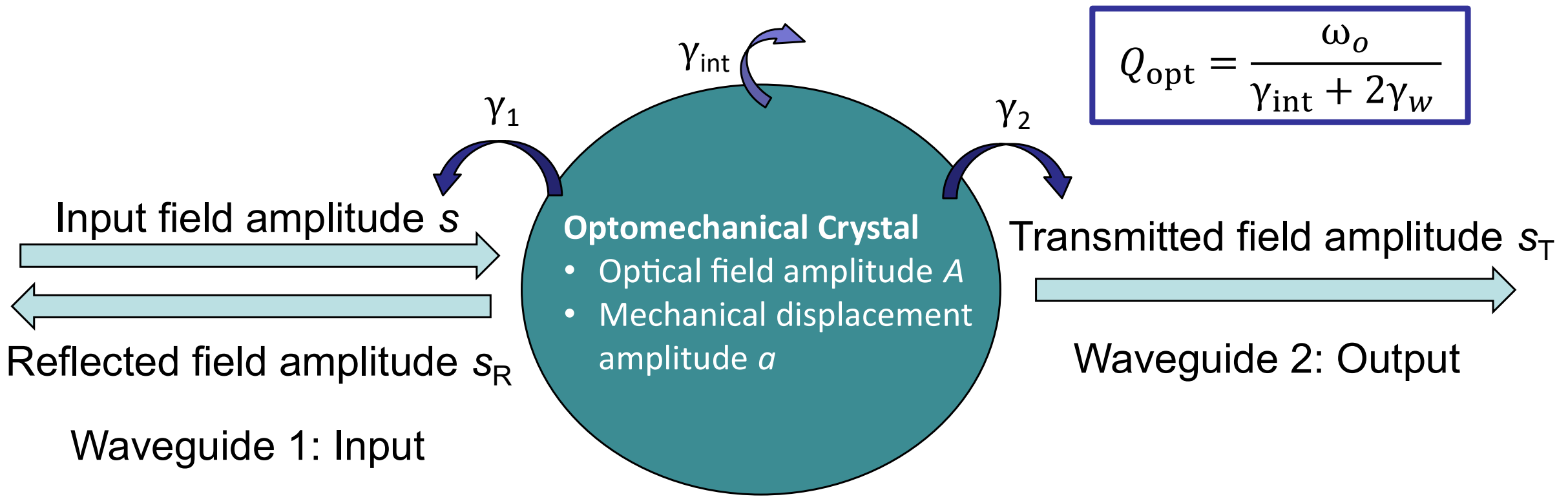

**Supplementary Figure 1. Optomechanical cavity with coupling waveguides.** A schematic of the optomechanical cavity edge-coupled to an input and output waveguide is shown. $\gamma_w$ is the waveguide coupling rate and $\gamma_{int}$ is the intrinsic optical loss rate. The optical resonance frequency is $\omega_0$ and $Q_{opt}$ is the optical quality factor.

Supplementary Fig. 1 shows a schematic of the optomechanical cavity that is coupled to an input and output waveguide. The cavity is designed such that it has an optical mode with a resonance frequency $\omega_0$ and optical field amplitude $A$. The mechanical mode has a frequency $\Omega_m$ and displacement amplitude $a$. There is an input optical field through waveguide 1 with incident amplitude $s$ normalized such that $|s|^2 \equiv P_{\mathrm{in}}$ is the optical power impinging on the cavity. In waveguide 1 there is also a reflected field with amplitude $s_R$. Waveguide 2 has a transmitted field with an amplitude $s_T$.

The OMC contains an optical cavity that has an intrinsic optical loss rate, $\gamma_{\mathrm{int}}$, and an extrinsic coupling rate between the optical cavity field and each waveguide of $\gamma_w$, resulting in a total optical quality factor $Q_{\mathrm{opt}} = \frac{\omega_0}{\gamma_{\mathrm{int}}+2\gamma_w}$. The acoustic mode in the cavity has an effective mass $m_{\mathrm{eff}} \equiv \rho V_m$ and total loss rate $\Gamma \equiv \Omega_m/Q_m$ where $\rho$ is the density, $V_m$ is the mechanical mode volume such that $m_{\mathrm{eff}} = \int dV\, \rho \left(\frac{|Q|}{\max(|Q|)}\right)^2$ where $Q$ is the displacement field, and $Q_m$ is the mechanical quality factor.

The optomechanical interaction results from the coupled optical and acoustic cavity. The optical resonant frequency depends on the mechanical displacement, and the mechanical mode is impacted by a radiation pressure force resulting from the optical field. The driving force is $F_{\mathrm{optical}}(t) = \pm |A(t)|^2/L_{\mathrm{OM}}$ where $L_{\mathrm{OM}}$ is an effective optomechanical length describing the strength of the optomechanical interaction such that $L_{\mathrm{OM}}^{-1} = \frac{g_{\mathrm{OM}}}{\omega_0}$ where $g_{\mathrm{OM}}$ is the optomechanical coupling factor. $g_{\mathrm{OM}}$ is related to the bare optomechanical coupling rate $g$ and mechanical oscillator zero-point fluctuation $x_{\mathrm{zpf}}$ as $g_{\mathrm{OM}} = \frac{g}{x_{\mathrm{zpf}}}$. We assume a sinusoidal mechanical displacement $a(t) = a_0 \sin(\Omega t)$ and that the optomechanical interaction is of linear order $\omega_0 \to \omega_0 \left(1 \mp \frac{a(t)}{L_{\mathrm{OM}}}\right)$. The coupled differential equations describing the optomechanical interaction are then [25]

$$\dot{A}(t) = \left(-\frac{\gamma}{2} - i\omega_0 \left(1 \mp \frac{a(t)}{L_{\mathrm{OM}}}\right)\right) A(t) + \sqrt{\gamma_w} s e^{-i\omega t} \tag{S1}$$

and

$$\ddot{a}(t) + \Gamma \dot{a}(t) + \Omega_m^2 a(t) = \pm \frac{|A(t)|^2}{m_{\mathrm{eff}} L_{\mathrm{OM}}}. \tag{S2}$$

We treat the acoustic mode as a small, time-dependent, and purely oscillatory perturbation that modulates the resonant frequency of the optical mode. Once the perturbed optical mode is solved for, the effect on the acoustic mode can be determined.

We seek the general solution for $A(t)$ given by $A_G(t) = A_H(t) + A_p(t)$ where $A_H(t)$ is the solution to the homogeneous differential equation and $A_p(t)$ is a particular solution, which has the form $A_p(t) = f(t) A_H(t)$ for a function $f(t)$ with a form that can be solved for from $\dot{A}(t)$ and $\dot{A}_H(t)$. The homogeneous differential equation for $A(t)$ is found by setting $s = 0$, which gives

$$\dot{A}_H(t) = \left(-\frac{\gamma}{2} - i\omega_0 + i\frac{\omega_0 a_0 \sin(\Omega t)}{L_{\mathrm{OM}}}\right) A_H(t) \tag{S3}$$

and has the solution

$$A_H(t) = A_0 \exp\left(\left(-\frac{\gamma}{2} - i\omega_0\right) t - i\frac{\omega_0 a_0 \cos(\Omega t)}{\Omega L_{\mathrm{OM}}}\right). \tag{S4}$$

To find the particular solution $A_p(t)$, we need to determine $f(t)$. We start with the relationship

$$\dot{A}(t) = \dot{A}_H(t) + \dot{A}_H(t) f(t) + A_H(t) \dot{f}(t) \tag{S5}$$

which gives

$$\dot{A}(t) = \left(-\frac{\gamma}{2} - i\omega_0 + i\frac{\omega_0 a_0 \sin(\Omega t)}{L_{\mathrm{OM}}}\right) [A_H(t) + A_H(t) f(t)] + \sqrt{\gamma_w} s e^{-i\omega t} \tag{S6}$$

where

$$\dot{f}(t) = \frac{\sqrt{\gamma_w} s e^{-i\omega t}}{A_H(t)} = s\sqrt{\gamma_w} e^{\left[\left(\frac{\gamma}{2} + i\omega_0 - i\omega\right) t + i\beta \cos(\Omega t)\right]}. \tag{S7}$$

The constant $A_0$ has been combined into $s$ and $\beta = \frac{\omega_0 a_0}{\Omega L_{\text{OM}}}$ is a modulation index. To determine $f(t)$ and therefore $A_p(t) = f(t)A_H(t)$, we must integrate the previous equation. The cosine part of the argument can be expanded into Bessel functions by the Jacobi-Anger expansion

$$\exp(\pm i\beta\cos(\Omega t)) = \sum_{n=-\infty}^{+\infty} (\pm i)^n J_n(\beta) e^{in\Omega t} \tag{S8}$$

where $J_n$ are Bessel functions of the first kind of index $n$. This results in the following for $f(t)$ after integrating:

$$f(t) = s\sqrt{\gamma_w} \sum_{n=-\infty}^{+\infty} \frac{i^n J_n(\beta)}{\left(\frac{\gamma}{2} + i(n\Omega - \Delta)\right)} e^{\left[\frac{\gamma}{2}+i(\omega_0-\omega+n\Omega)\right]t}. \tag{S9}$$

where $\Delta = \omega - \omega_0$ is the detuning. Therefore,

$$A_p(t) = s\sqrt{\gamma_w} \sum_{n=-\infty}^{+\infty} \frac{i^n J_n(\beta)}{\left(\frac{\gamma}{2} + i(n\Omega - \Delta)\right)} e^{i[(n\Omega-\omega)t-\beta\cos(\Omega t)]}. \tag{S10}$$

The general solution is $A_G(t) = A_H(t) + A_p(t)$. However, $A_H(t)$ is exponentially damped at rate $\frac{\gamma}{2}$ and therefore the general solution converges to the steady-state solution $A_p(t)$.

The power exiting the cavity is

$$|s_T|^2 = \gamma_w \left|A_p(t)\right|^2 \tag{S11}$$

where $s_T$ is the transmitted field amplitude and

$$\left|A_p(t)\right|^2 = |s|^2\gamma_w \sum_{n,m} \frac{i^{(n-m)} J_n(\beta) J_m(\beta)}{\left(\frac{\gamma}{2} + i(n\Omega - \Delta)\right)\left(\frac{\gamma}{2} - i(m\Omega - \Delta)\right)} e^{i(n-m)\Omega t}. \tag{S12}$$

The transmission is then

$$T = \frac{|s_T|^2}{|s|^2} = \gamma_w^2 \sum_{n,m} \frac{i^{(n-m)} J_n(\beta) J_m(\beta)}{\left(\frac{\gamma}{2} + i(n\Omega - \Delta)\right)\left(\frac{\gamma}{2} - i(m\Omega - \Delta)\right)} e^{i(n-m)\Omega t}. \tag{S13}$$

The effect of the acoustic mode on the optical mode is a set of sidebands that are generated in the cavity by the optomechanical interaction. These sidebands can be observed by measurement of a radio frequency (RF) spectrum of the transmitted power $|s_T|^2$ through photomixing on a square-law optical detector. In the limit of small optical modulation ($\beta \ll 1$) we only need to consider the first order sidebands given by the products $J_0(\beta)J_{\pm1}(\beta)$ [54]. The terms where $n = m = 0$ gives a DC signal only. Therefore, we only have four terms remaining where $n = 0, m = \pm 1$ and $n = \pm 1, m = 0$. In addition, for $\beta \to 0$ we can use the approximation $J_n(\beta \to 0) \approx \frac{1}{n!}\left(\frac{\beta}{2}\right)^n$ and therefore $J_0(\beta) \approx 1$ and $J_{\pm1}(\beta) \approx \pm {}^{\beta}/_{2}$. The power oscillating at $\Omega$ is then

$$T_{\Omega} = \gamma_w^2 \left[ \frac{i^{-1} J_0(\beta) J_1(\beta) e^{-i\Omega t}}{\left(\frac{\gamma}{2} + i(-\Delta)\right)\left(\frac{\gamma}{2} - i(\Omega - \Delta)\right)} + \frac{i J_0(\beta) J_{-1}(\beta) e^{i\Omega t}}{\left(\frac{\gamma}{2} + i(-\Delta)\right)\left(\frac{\gamma}{2} - i(-\Omega - \Delta)\right)} \right.$$
$$\left. + \frac{i J_1(\beta) J_0(\beta) e^{i\Omega t}}{\left(\frac{\gamma}{2} + i(\Omega - \Delta)\right)\left(\frac{\gamma}{2} - i(-\Delta)\right)} + \frac{i^{-1} J_{-1}(\beta) J_0(\beta) e^{-i\Omega t}}{\left(\frac{\gamma}{2} + i(-\Omega - \Delta)\right)\left(\frac{\gamma}{2} - i(-\Delta)\right)} \right]. \quad \text{(S14)}$$

Using the fact that $i^{-1} = 1/i = -i$ and $J(\beta)_{-1} = -J_1(\beta)$:

$$T_{\Omega} = \gamma_w^2 \left[ \frac{-i J_0(\beta) J_1(\beta) e^{-i\Omega t}}{\left(\frac{\gamma}{2} - i\Delta\right)\left(\frac{\gamma}{2} - i(\Omega - \Delta)\right)} - \frac{i J_0(\beta) J_1(\beta) e^{i\Omega t}}{\left(\frac{\gamma}{2} - i\Delta\right)\left(\frac{\gamma}{2} + i(\Omega + \Delta)\right)} \right.$$
$$\left. + \frac{i J_1(\beta) J_0(\beta) e^{i\Omega t}}{\left(\frac{\gamma}{2} + i(\Omega - \Delta)\right)\left(\frac{\gamma}{2} + i\Delta\right)} + \frac{i J_1(\beta) J_0(\beta) e^{-i\Omega t}}{\left(\frac{\gamma}{2} - i(\Omega + \Delta)\right)\left(\frac{\gamma}{2} + i\Delta\right)} \right]. \quad \text{(S15)}$$

The four terms can be written as two terms plus their complex conjugates:

$$T_{\Omega} = J_0(\beta) J_1(\beta) \gamma_w^2 \left[ \left( \frac{-i e^{-i\Omega t}}{\left(\frac{\gamma}{2} - i\Delta\right)\left(\frac{\gamma}{2} - i(\Omega - \Delta)\right)} + c.c. \right) \right.$$
$$\left. + \left( \frac{-i e^{i\Omega t}}{\left(\frac{\gamma}{2} - i\Delta\right)\left(\frac{\gamma}{2} + i(\Omega + \Delta)\right)} + c.c. \right) \right] \quad \text{(S16)}$$

and using the property of the complex conjugate that the sum of a complex number and its complex conjugate is equal to twice the real part of the complex number we have the following:

$$T_{\Omega} = 2 J_0(\beta) J_1(\beta) \gamma_w^2 \left[ \mathrm{Re} \left\{ \frac{-i e^{-i\Omega t}}{\left(\frac{\gamma}{2} - i\Delta\right)\left(\frac{\gamma}{2} - i(\Omega - \Delta)\right)} + \frac{-i e^{i\Omega t}}{\left(\frac{\gamma}{2} - i\Delta\right)\left(\frac{\gamma}{2} + i(\Omega + \Delta)\right)} \right\} \right]. \quad \text{(S17)}$$

We now will seek the expression for $T_{\Omega}$ that breaks the spectrum into two components: in phase and in quadrature with the motion of the mechanical oscillator. One is proportional to $\sin(\Omega t)$ and the other is proportional to $\cos(\Omega t)$ such that:

$$T_{\Omega} = \beta \gamma_w^2 \left[ \cos(\Omega t) \left( \frac{2\gamma}{\gamma^2 + 4\Delta^2} \right) \left( \frac{\Omega}{\left(\frac{\gamma}{2}\right)^2 + (\Omega - \Delta)^2} - \frac{\Omega}{\left(\frac{\gamma}{2}\right)^2 + (\Omega + \Delta)^2} \right) \right.$$
$$\left. + \sin(\Omega t) \left( \frac{4\Omega}{\gamma^2 + 4\Delta^2} \right) \left( \frac{\Omega - \Delta}{\left(\frac{\gamma}{2}\right)^2 + (\Omega - \Delta)^2} - \frac{\Omega + \Delta}{\left(\frac{\gamma}{2}\right)^2 + (\Omega + \Delta)^2} \right) \right]. \quad \text{(S18)}$$

where we have used $J_0(\beta) \approx 1$ and $J_{\pm 1}(\beta) \approx \pm {}^{\beta}/_{2}$.

We have expressed the power oscillating at $\Omega$ in terms of

$$T_\Omega = A_c \cos(\Omega t) + A_s \sin(\Omega t) \tag{S19}$$

such that $A_c$ is the in-quadrature component and $A_s$ is the in-phase component. The output power at $\Omega$ is

$$P_{\mathrm{out},\Omega} = P_{\mathrm{in}} \sqrt{A_c^2 + A_s^2}. \tag{S20}$$

We now will look at the effect of the optical mode on the acoustic mode. This analysis allows us to ultimately predict the mechanical output power for an optomechanical ultrasound transmitter. The optical force is given by

$$F_{\mathrm{opt}}(t) = \frac{|A_p(t)|^2}{L_{\mathrm{OM}}} = \frac{|s|^2 \gamma_w}{L_{\mathrm{OM}}} \sum_{n,m} \frac{i^{(n-m)} J_n(\beta) J_m(\beta)}{\left(\frac{\gamma}{2} + i(n\Omega - \Delta)\right)\left(\frac{\gamma}{2} - i(m\Omega - \Delta)\right)} e^{i(n-m)\Omega t}. \tag{S21}$$

In the limit of small optical modulation ($\beta \ll 1$), the only DC component occurs when $n = m = 0$. The DC component of $F_{\mathrm{opt}}$ is then

$$F_{\mathrm{opt,DC}} = \frac{|s|^2 \gamma_w}{L_{\mathrm{OM}}} \frac{1}{\frac{\gamma^2}{4} + \Delta^2} = \frac{P_{\mathrm{in}}}{L_{\mathrm{OM}}} \frac{4\left({}^{\gamma_w}/_{\gamma^2}\right)}{1 + 4\left(\frac{\Delta}{\gamma}\right)^2}. \tag{S22}$$

For the RF optical force $F_{\mathrm{opt},\Omega} = \frac{T_\Omega |s|^2}{L_{\mathrm{OM}} \gamma_w}$, we can use the previous result for $T_\Omega$ because $|A_p(t)|^2 = \frac{T_\Omega |s|^2}{\gamma_w}$:

$$\begin{aligned} F_{\mathrm{optical},\Omega} = \frac{\beta \gamma_w |s|^2}{L_{\mathrm{OM}}} \Bigg[ & \cos(\Omega t) \left(\frac{2\gamma}{\gamma^2 + 4\Delta^2}\right) \left( \frac{\Omega}{\left(\frac{\gamma}{2}\right)^2 + (\Omega - \Delta)^2} - \frac{\Omega}{\left(\frac{\gamma}{2}\right)^2 + (\Omega + \Delta)^2} \right) \\ & + \sin(\Omega t) \left(\frac{4\Omega}{\gamma^2 + 4\Delta^2}\right) \left( \frac{\Omega - \Delta}{\left(\frac{\gamma}{2}\right)^2 + (\Omega - \Delta)^2} - \frac{\Omega + \Delta}{\left(\frac{\gamma}{2}\right)^2 + (\Omega + \Delta)^2} \right) \Bigg]. \end{aligned} \tag{S23}$$

The RF force is then written in terms of an in-quadrature component $F_c$ and in in-phase component $F_s$ such that

$$F_{\mathrm{opt},\Omega} = F_c \cos(\Omega t) + F_s \sin(\Omega t) = F_c \frac{\dot{a}(t)}{\Omega a_0} + F_s \frac{a(t)}{a_0}. \tag{S24}$$

Now that we have an expression for $F_{\mathrm{opt}}$ that includes optomechanical coupling effects, we can examine the effect of the optical mode on the mechanical mode. Beginning with the equation of motion for the mechanical oscillator:

$$\ddot{a}(t) + \Gamma\dot{a}(t) + \Omega_m^2 a(t) = \frac{F_{\mathrm{opt}}}{m_{\mathrm{eff}}} = \frac{1}{m_{\mathrm{eff}}}\left(F_c\frac{\dot{a}(t)}{\Omega a_0} + F_s\frac{a(t)}{a_0} + F_{\mathrm{opt,DC}}\right) \tag{S25}$$

The impact of $F_c$ and $F_s$ on the mechanical oscillator is then to modify the mechanical damping rate and the mechanical frequency, respectively, such that the modified equation of motion is

$$\ddot{a}(t) + (\Gamma + \Gamma_{\mathrm{OM}})\dot{a}(t) + (\Omega_m^2 + 2\Omega_m\delta\Omega)a(t) = \frac{F_{\mathrm{opt,DC}}}{m_{\mathrm{eff}}} \tag{S26}$$

where $F_{\mathrm{opt,DC}}$ is the optical DC force,

$$\Gamma_{\mathrm{OM}} = -\frac{2\omega_0\gamma_w|s|^2}{m_{\mathrm{eff}}\Omega L_{\mathrm{OM}}^2}\left(\frac{1}{\gamma^2 + 4\Delta^2}\right)\left(\frac{\gamma}{\left(\frac{\gamma}{2}\right)^2 + (\Omega - \Delta)^2} - \frac{\gamma}{\left(\frac{\gamma}{2}\right)^2 + (\Omega + \Delta)^2}\right) \tag{S27}$$

is the optomechanical damping rate, and

$$\delta\Omega = -\frac{2\omega_0\gamma_w|s|^2}{m_{\mathrm{eff}}\Omega_m L_{\mathrm{OM}}^2}\left(\frac{1}{\gamma^2 + 4\Delta^2}\right)\left(\frac{\Omega - \Delta}{\left(\frac{\gamma}{2}\right)^2 + (\Omega - \Delta)^2} - \frac{\Omega + \Delta}{\left(\frac{\gamma}{2}\right)^2 + (\Omega + \Delta)^2}\right) \tag{S28}$$

is the optomechanically induced change in the mechanical resonant frequency. The optomechanically induced change in the mechanical resonant frequency is often referred to as the optical spring effect.

**Supplementary Note 2. Optical cavity loss rates**

One configuration for light delivery to the optical cavity is through edge-coupled waveguides, which minimizes interference with phonon beam generation and propagation in the final structure. A model of the electric field distribution in the optical cavity with the labeled field amplitudes and optical loss rates is shown in Supplementary Fig. 2(a). In the input waveguide, there is an input optical signal with a field amplitude $s$ and a reflected optical signal with an amplitude $s_R$. In the output waveguide, there is a transmitted field with amplitude $s_T$. Scattering and radiation losses at the waveguide-mirror interface are reduced with a linear taper of the mirror. We can tailor the optical loss rates by modifying the number of mirror unit cells surrounding the optical cavity. Supplementary Fig. 2(b) shows a plot of the modeled $Q_{\mathrm{opt}}$, $Q_{\mathrm{int}}$, and $Q_w$ as a function of the number of mirror unit cells where $Q_{\mathrm{opt}} = \left(\frac{1}{Q_{\mathrm{int}}} + \frac{1}{Q_w}\right)^{-1}$. $Q_{\mathrm{opt}} = \omega_0/\gamma$ is the total optical quality factor, $Q_{\mathrm{int}} = \omega_0/\gamma_{\mathrm{int}}$ is the intrinsic quality factor from radiative and absorption losses, and $Q_w = \omega_0/2\gamma_w$ is the waveguide-coupled optical quality factor. The optical cavity loss rate is plotted a function of the number of mirror unit cells in Supplementary Fig. 2(c). We design the cavity to operate in a strongly overcoupled regime where the total optical loss rate is dominated by coupling to the two waveguides. With 5 mirror unit cells, the modeled total optical loss rate is approximately equal to the total waveguide-coupled loss rate, and $Q_{\mathrm{opt}} = 1.5 \times 10^5$.

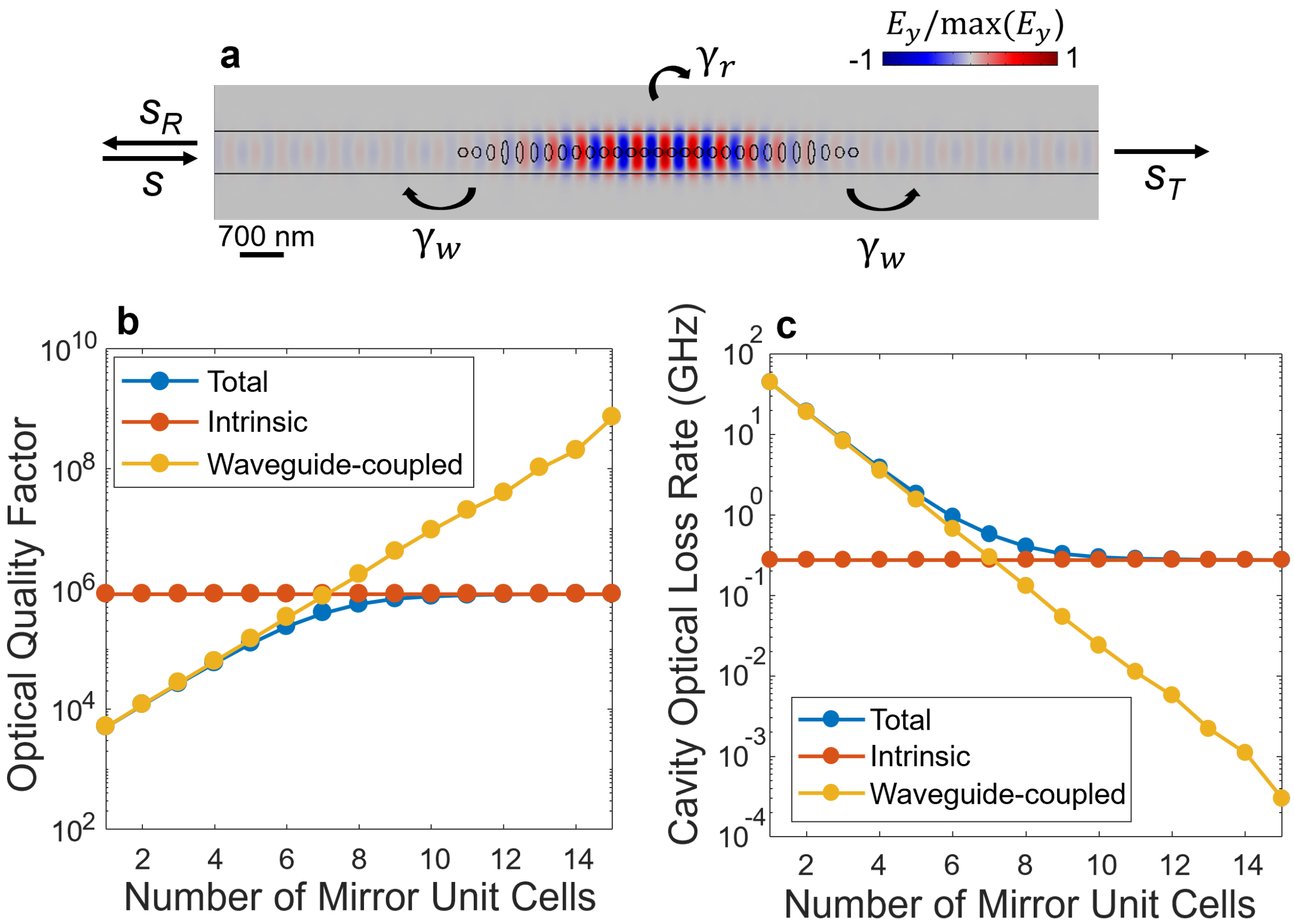

**Supplementary Figure 2. Optical loss rates of the OMC.** (a) A model of the electric field distribution on-resonance in the optical cavity with labeled optical field amplitudes and optical loss rates. The modeled (b) optical quality factor and (c) optical loss rates as a function of the number of mirror unit cells.

**Supplementary Note 3. Analysis of phononic bandstructure in water**

When attempting to compute the phononic bandstructure in water using a viscous fluid model, we found that the eigenmode simulation in the presence of water results in a large spectral density of spurious modes. Therefore, we instead assess if the expected stop band is preserved when the phononic mirror is immersed in water by FEM modeling of an array of phononic mirror unit cells in a viscous fluid with an input mechanical power introduced by a mechanical forcing function. We compute the output mechanical power as the frequency of the input forcing function is swept across the expected stop band. A plot of the ratio of the mechanical output power to the mechanical input power as a function of the frequency of the forcing function for an increasing number of phononic mirror unit cells is shown in Supplementary Fig. 3. As can be seen, the mechanical stop band is preserved when immersed in water.

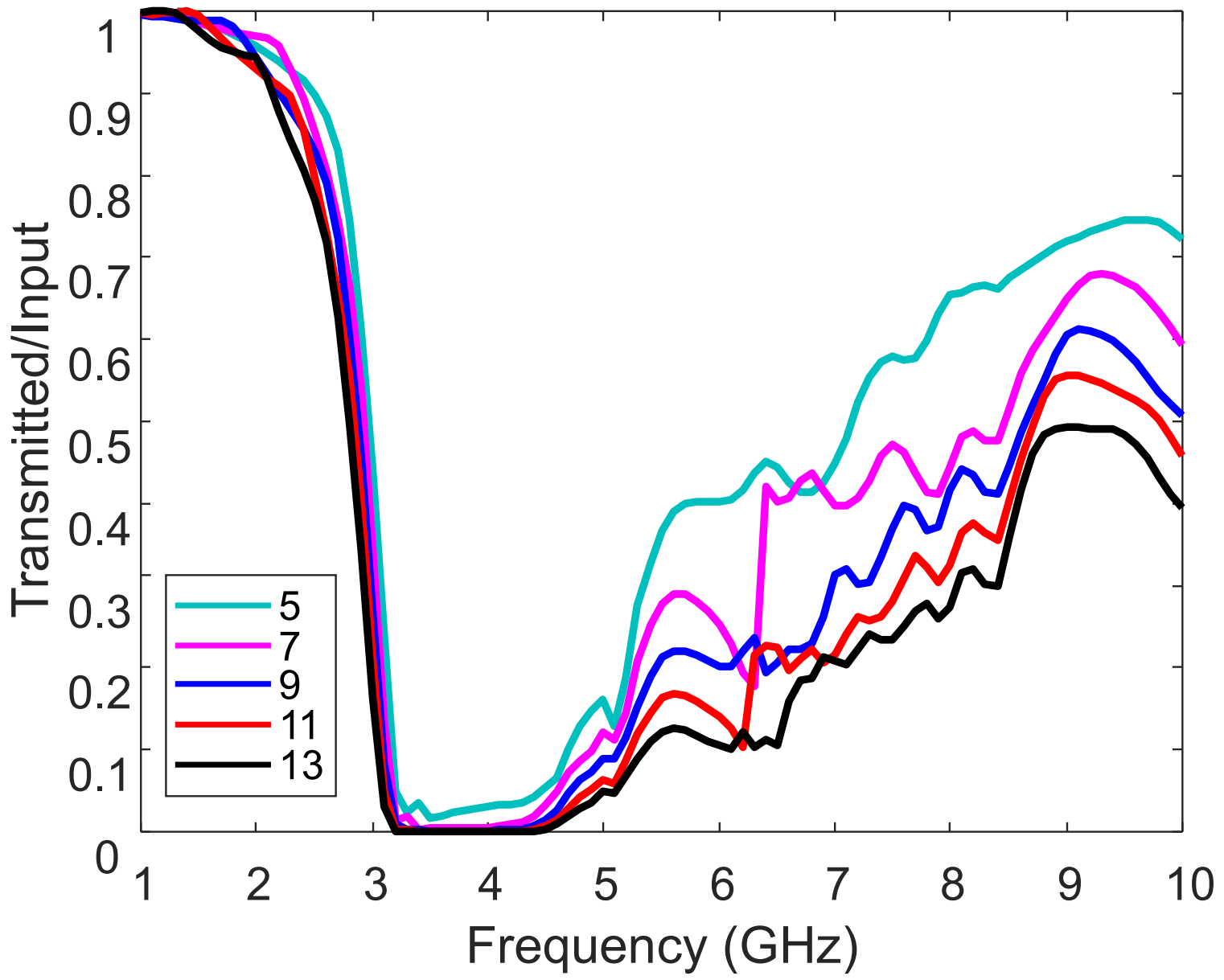

**Supplementary Figure 3. Phononic mirror in water.** A plot of the ratio between the transmitted mechanical power and the input mechanical power to an array of mirror unit cells as a function of frequency for the mechanical forcing function from a 3D FEM acoustic-structure interaction study with a viscous fluid model. The legend indicates the number of mirror unit cells in the array.

**Supplementary Note 4. Optomechanical length and mechanical oscillator effective mass**

The effective optomechanical length is given by $L_{\mathrm{OM}}^{-1} = \frac{1}{\omega_0}\frac{d\omega_0}{da} = \frac{g_{\mathrm{OM}}}{\omega_0}$ where $\frac{d\omega_0}{da}$ quantifies the change in the optical resonance frequency, $\omega_0$, with a change in mechanical displacement, $a$, and $g_{\mathrm{OM}}$ is the optomechanical coupling factor. $g_{\mathrm{OM}}$ is related to the bare optomechanical coupling rate, $g$, and mechanical oscillator zero-point fluctuation, $x_{\mathrm{zpf}}$, by $g_{\mathrm{OM}} = g/x_{\mathrm{zpf}}$. The rate $g$ measures the frequency shift in the optical mode frequency imparted by the zero-point motion of the mechanical oscillator, $x_{\mathrm{zpf}}$, and is given by

$$g = \frac{d\omega_0}{da} x_{\mathrm{zpf}} = \frac{d\omega_0}{da}\sqrt{\frac{\hbar}{2m_{\mathrm{eff}}\Omega_m}} \tag{S29}$$

where $\hbar$ is Planck's constant divided by 2π, $m_{\mathrm{eff}}$ is the effective mass, and $\Omega_m$ is the mechanical resonance frequency.

To determine the optomechanical coupling rate, we adopt a previously established derivation [40] that considers a combination of a dielectric moving boundary effect and the photoelastic effect such that

$$\frac{d\omega_0}{da} \approx \frac{d\omega_0}{da}_{\mathrm{MB}} + \frac{d\omega_0}{da}_{\mathrm{PE}} \tag{S30}$$

where $\frac{d\omega_0}{da}_{\text{MB}}$ is the optical frequency shift due to the moving boundary effect and $\frac{d\omega_0}{da}_{\text{PE}}$ is the optical frequency shift due to the photoelastic effect. The photoelastic coefficients for silicon have been measured previously at optical wavelengths of 1.15 µm and 3.39 µm [55], but not at 1.3 µm, and therefore we interpolate the available experimental data and use photoelastic coefficients of $p_{11}$=-0.1006, $p_{12}$ =0.0096, and $p_{44}$=-0.051. A 3D-FEM model of the optomechanical crystal interacting with a viscous fluid was used to assess the mechanical frequency, effective mass, optomechanical coupling rate, and optomechanical length. The resulting values are listed in Supplementary Table 1.

Supplementary Table 1: Modeled optomechanical parameters.

| Parameter | Description | Value |
|---|---|---|
| $m_{\text{eff}}$ | Effective mass | 307 fg |
| $g_{\text{MB}}$ | Optomechanical coupling rate due to dielectric moving boundary | $2\pi \times$ -138 kHz |
| $g_{\text{PE}}$ | Optomechanical coupling rate due to photoelastic effect | $2\pi \times$ 666 kHz |
| $g$ | Total optomechanical coupling rate | $2\pi \times$ 528 kHz |
| $\omega_0$ | Optical angular frequency | $2\pi \times$ 230 THz |
| $\Omega_m$ | Mechanical angular frequency | $2\pi \times$ 4.6 GHz |
| $L_{\text{OM}}$ | Optomechanical length | 1 µm |

**Supplementary Note 5. Steady-state mechanical displacement for regenerative optomechanical oscillations**

When $P_{in} > P_t$, the overall damping rate $\Gamma + \Gamma_{OM}$ becomes negative, causing thermal fluctuations to grow until the mechanical oscillation amplitude reaches a steady-state value, $a_s$, where the nonlinear optomechanical interaction restores the net damping to zero. Following established treatments of regenerative optomechanical oscillations, we assume a power-independent optical linewidth in the idealized linear cavity limit.[37,38] The effective mechanical linewidth, $\Gamma_{eff}$, is given by [56]

$$\begin{aligned} \Gamma_{\text{eff}} &= \Gamma + \Gamma_{\text{OM}} & \qquad P_{\text{in}} < P_t \\ \Gamma_{\text{eff}} &= \frac{\Gamma^2 k_B T}{2P_{\text{out}}}, & \qquad P_{\text{in}} > P_t \end{aligned} \tag{S31}$$

where $P_{out}$ is the mechanical output power. Equation (S31) is equivalent to the Schawlow-Townes limit for the linewidth, [57] given that thermomechanical noise is the dominant noise source. $P_{out}$ is determined by $P_{out} = U_m\Gamma = \frac{1}{2} m_{eff}\Omega_m^2 a_0^2 \Gamma$ where $U_m$ is the total stored mechanical energy [56,58,59]. The stable values of $a_0 = a_s$ occur when the energy lost to friction in one mechanical cycle is balanced by the energy gained due to the radiation pressure force [38,60] and can be found by following an established procedure [60]. We first assume that the self-induced mechanical oscillations take the form $a(t) = a_s sin(\Omega t)$ and then by examining the balance between the time-averaged mechanical power due to the radiation pressure force

$$P_{\mathrm{opt}} = \langle F_{\mathrm{opt}}\dot{a}\rangle = \frac{1}{L_{\mathrm{OM}}}\langle |A(t)|^2 \dot{a}\rangle \tag{S32}$$

and the mechanical power lost to friction

$$P_{\mathrm{friction}} = m_{\mathrm{eff}}\Gamma\langle \dot{a}^2\rangle. \tag{S33}$$

The expression for $A(t)$ is given by

$$A(t) = s\sqrt{\gamma_w}e^{i\varphi(t)}\sum_n i^n \frac{J_n(\beta)}{\frac{\gamma}{2} + i(n\Omega - \Delta)}e^{in\Omega t} \tag{S34}$$

where $\varphi(t)$ is a global phase given by $\varphi(t) = \beta\cos(\Omega t)$. We then examine the ratio

$$\frac{P_{\mathrm{opt}}}{P_{\mathrm{friction}}} = \frac{\langle F_{\mathrm{opt}}\dot{a}\rangle}{m_{\mathrm{eff}}\Gamma\langle \dot{a}^2\rangle} \tag{S35}$$

as a function of $\beta$ and $\Delta$. Self-induced oscillations occur for values of $\beta$ where $P_{\mathrm{opt}} = P_{\mathrm{friction}}$ and these solutions are stable if $P_{\mathrm{opt}}/P_{\mathrm{friction}}$ decreases with increasing $\beta$.

**Supplementary Note 6. Thermal analysis of the OMC transmitter**

Modeling nonlinear optical effects that lead to self-pulsing in the OMC transmitter is done through a system of coupled equations for the optical mode amplitude, $A$, free-carrier density, $N$, and the temperature increase relative to the ambient temperature, $\Delta T$,[50]

$$\dot{A}(t) = \left(-\frac{\gamma_t}{2} + i\Delta\omega(t)\right)A(t) + \sqrt{\gamma_w}s \tag{S36}$$

$$\dot{N}(t) = -\gamma_{\mathrm{fc}}N(t) + \alpha_{\mathrm{fc}}U(t)^2 \tag{S37}$$

$$\dot{\Delta T}(t) = -\gamma_{\mathrm{th}}\Delta T(t) + \alpha_{\mathrm{th}}P_{\mathrm{abs}}(t) \tag{S38}$$

where $\gamma_t = \gamma_{int} + 2\gamma_w + \gamma_{lin} + \gamma_{FCA} + \gamma_{TPA}$ is the total optical loss rate, $\gamma_{lin}$ is the loss due to linear absorption, $\gamma_{FCA}$ is the free-carrier absorption (FCA), $\gamma_{TPA}$ is the two-photon absorption (TPA), $\Delta\omega$ is the detuning of the input light frequency, $\gamma_{fc}$ is the free carrier recombination rate, $\alpha_{fc} = \frac{\beta_{Si}c^2}{2\hbar\omega_l n_g^2 V_{TPA}{}^2}$ quantifies the carrier generation where $\beta_{Si}$ is the TPA coefficient in silicon, $c$ is the speed of light, $\hbar$ is Planck's constant, $\omega_l$ is the laser frequency, $n_g$ is the group refractive index, $V_{TPA}$ is the effective mode volume for TPA, $U(t) = |A(t)|^2$ is the stored optical energy in the cavity, $\gamma_{th}$ is the thermal decay rate, $\alpha_{th}$ is the thermal absorption coefficient, and $P_{abs}(t) = \gamma_a U(t)$ is the absorbed optical power in the silicon where $\gamma_a = \gamma_{lin} + \gamma_{TPA} + \gamma_{FCA}$ is the optical loss rate due to absorption in the silicon. The thermal absorption coefficient, $\alpha_{th}$, is given by $\alpha_{th} = \frac{1}{\rho_{Si}c_{p,Si}V_{th}}$ where $\rho_{Si}$ is the density of silicon, $c_{p,Si}$ is the constant-pressure specific heat capacity of silicon, $V_{th}$ is the thermal mode volume, $\gamma_{TPA}$ is found according to $\gamma_{TPA}(t) = \frac{\beta_{Si}c^2}{n_g^2 V_{TPA}}U(t)$, and $\gamma_{FCA}$ is found according to $\gamma_{FCA}(t) = \frac{\sigma_{Si}c}{n_g}N(t)$ where $\sigma_{Si}$ is the FCA cross-section in silicon. The total optical detuning, $\Delta\omega$, is given by

$$\Delta\omega(t) = \Delta\omega_i + \frac{\omega_0}{n_{\mathrm{Si}}}\left(\frac{dn_{\mathrm{Si}}}{dT}\Delta T(t) + \frac{dn_{\mathrm{Si}}}{dN}N(t)\right) \tag{S39}$$

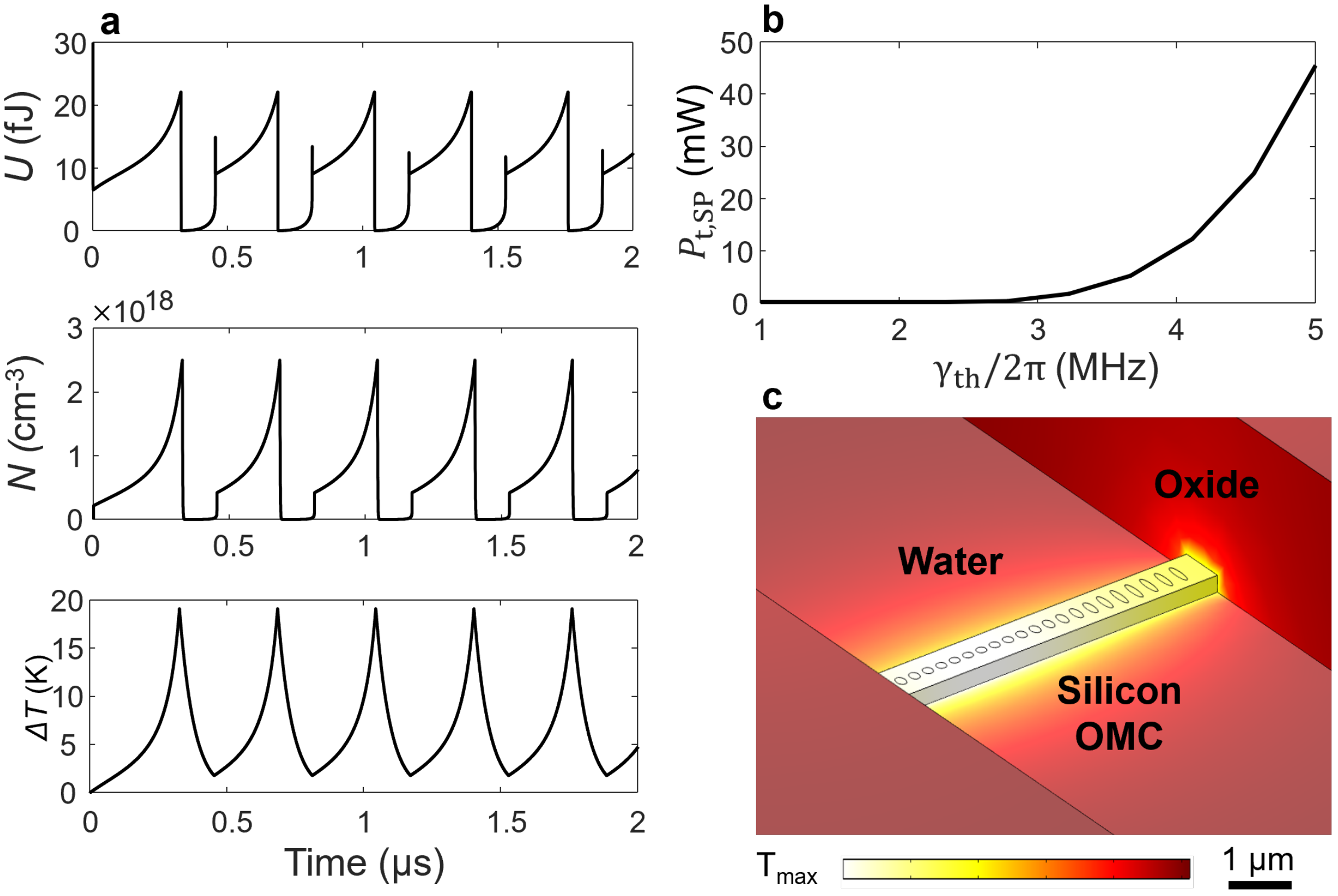

**Supplementary Figure 4. Thermal properties of the OMC transmitter.** (a) $U(t)$, $N(t)$, and $\Delta T(t)$ for an OMC with $\gamma_{th}/2\pi = 3$ MHz corresponding to heat dissipation only through thermal conduction in the silicon. (b) $P_{t,SP}$ as a function of increasing $\gamma_{th}/2\pi$. (c) Steady-state temperature distribution in the silicon OMC submerged in water where heat transfer can occur through thermal conduction and convection in the surrounding aqueous solution.

where $\Delta\omega_i$ is the initial detuning, $\frac{dn_{Si}}{dT}$ is the thermo-optic coefficient of silicon and $\frac{dn_{Si}}{dN}$ is a coefficient that gives the change in refractive index of silicon with a change in the free carrier density.

Equation (S37) then describes the balance between free-carrier generation from TPA and free-carrier recombination according to $\gamma_{\mathrm{fc}}$ while Eqn. (S38) describes the balance between heat generated from free carrier absorption and the heat dissipated according to $\gamma_{\mathrm{th}}$. The value for $\gamma_{\mathrm{fc}}$ is approximated according to reported values [61]. A FEM heat transfer model of our OMC ultrasound device with edge-coupled oxide-clad silicon waveguides on a silicon substrate was built to obtain the value for $\gamma_{\mathrm{th}}$. An initial $\Delta T$ was modeled along the length of the beam in a Gaussian distribution such that $\Delta T$=2.5 K at the center of the beam. The maximum temperature in the beam was recorded with respect to time and then fit to an exponential decay equation to extract $\gamma_{\mathrm{th}}$. Modeling thermal conduction in the silicon and silicon dioxide and convection into the air via a heat transfer coefficient of 5 W/m$^2$-K, $\gamma_{\mathrm{th}}/2\pi = 3 \pm 1$ MHz where the error corresponds to plus/minus one standard deviation from the fitting. Equations (S36-S38) can be solved through numerical integration to find $U(t)$, $N(t)$, and $\Delta T(t)$ for the set of parameters relevant to the transmitter given in Supplementary Table 2. Supplementary Fig. 4(a) shows $U(t)$, $N(t)$, and $\Delta T(t)$ for $P_{\text{in}} = 15$ mW and $\gamma_{\text{th}}/2\pi = 3$ MHz. Self-pulsing occurs due to the change in $U(t)$ driven by changing $N(t)$ and $\Delta T(t)$.

However, this value for $\gamma_{th}$ does not capture the thermal dissipation when the transmitter is submerged in an aqueous solution. When the water is included in the FEM model, $\gamma_{th}$ increases to $\gamma_{th}/2\pi = 40 \pm 8$ MHz likely due to heat transfer through thermal conduction and convection in the surrounding aqueous solution. As shown in the plot of $P_{t,SP}$ as a function of $\gamma_{th}$ in Supplementary Fig. 4(b), $P_{t,SP}$ increases rapidly with $\gamma_{th}$. The faster thermal relaxation removes heat before a large thermo-optic detuning can accumulate. The steady-state temperature rise scales according to $P_{abs}/\gamma_{th}$ so that a larger $\gamma_{th}$ requires higher absorbed power to produce the same resonance shift. In the full nonlinear cavity model, this baseline scaling is amplified by the detuning-dependent intracavity energy, nonlinear absorption, free-carrier generation, and power-dependent linewidth broadening. When the water is included in the FEM model (Supplementary Fig. 4(c)), the computed temperature decay rate increases to $\gamma_{th}/2\pi = 40 \pm 8$ MHz. The modeling results suggest that the increased thermal decay rate substantially raises the self-pulsing threshold, thereby extending the optical power range over which thermal self-pulsing is suppressed.

Supplementary Table 2: Parameters for thermal modeling of the OMC transmitter.

| Parameter | Description | Value | Source |
|---|---|---|---|
| $\lvert s\rvert^2$ | Optical input power | Varies | Model |
| $\gamma_{\text{fc}}$ | Free-carrier recombination rate | $10^9$ s$^{-1}$ | [61] |
| $\gamma_{\text{th}}$ | Thermal decay rate | Varies | Model |
| $\beta_{Si}$ | TPA coefficient | $7.5 \times 10^{-12}$ m · W$^{-1}$ | [62] |
| $\omega_0$ | Optical angular frequency | $2\pi \times 231$ THz | Model |
| $n_g$ | Group index | 3.88 | Model |
| $n_{\text{Si}}$ | Silicon refractive index | 3.503 | [63] |
| $\gamma_{\text{lin}}$ | Linear absorption | ~$2\pi \times 55$ kHz | [63] |
| $\lambda_0$ | Optical resonance wavelength | 1.3 μm | Model |
| $V_0$ | Optical mode volume | $1.3\ (\lambda_0/n_{\text{Si}})^3$ | Model |
| $V_{\text{TPA}}$ | TPA effective mode volume | $3.3\ (\lambda_0/n_{\text{Si}})^3$ | Model |
| $V_{\text{th}}$ | Thermal mode volume | $2\times10^{-18}$ m$^3$ | Model |
| $\rho_{Si}$ | Density of silicon | 2330 kg · m$^{-3}$ | [64] |
| $c_{p,Si}$ | Silicon specific heat capacity | 700 J · kg$^{-1}$ · K$^{-1}$ | [64] |
| $\sigma_{Si}$ | Free carrier absorption cross-section | $5.12 \times 10^{-22}$ m$^2$ | [65] |
| $\Delta\omega_i$ | Initial detuning | $2\pi \times 4.6$ GHz | Model |
| $dn_{\text{Si}}/dT$ | Silicon thermo-optic coefficient | $1.94 \times 10^{-4}$ K$^{-1}$ | [66] |
| $dn_{\text{Si}}/dN$ | Refractive index change with changing free carrier density | $-1.5 \times 10^{-27}$ m$^3$ | [67] |

**Supplementary Note 7. Noise equivalent pressure derivation**

The total noise equivalent pressure (NEP) for the optomechanical ultrasound receiver is calculated by considering the relevant distinct noise mechanisms according to

$$\mathrm{NEP} = \sqrt{\mathrm{p}_{\mathrm{a}}^2 + \mathrm{p}_{\mathrm{th}}^2 + \mathrm{p}_{\mathrm{BA}}^2 + \mathrm{p}_{\mathrm{TR}}^2 + \mathrm{p}_{\mathrm{TM}}^2 + \mathrm{p}_{\mathrm{SN}}^2 + \mathrm{p}_{\mathrm{det}}^2 + \mathrm{p}_{\mathrm{amp}}^2} \tag{S40}$$

where $\mathrm{p_a}$ is the NEP associated with the thermal-acoustic noise of the medium, $\mathrm{p_{th}}$ is the NEP associated with the mechanical oscillator's intrinsic thermal noise, $\mathrm{p_{BA}}$ is the NEP due to back-action noise, $\mathrm{p_{TR}}$ is the thermo-refractive noise, $\mathrm{p_{TM}}$ is the thermal-mechanical noise, $\mathrm{p_{SN}}$ is the NEP due to shot noise, $\mathrm{p_{det}}$ is the NEP from detector noise, and $\mathrm{p_{amp}}$ is the NEP due to the RF amplifier with noise factor $F_n$.

The NEP due to thermal noise is due to two independent contributions: one due to the thermal fluctuations intrinsic to the mechanical oscillator, characterized by an internal dissipation $\Gamma_{\mathrm{int}}$, and the other due to the fluid-induced noise from interaction of the oscillator with the fluid, characterized by the loss rate $\Gamma_f$. The total mechanical damping rate is $\Gamma = \Gamma_{\mathrm{int}} + \Gamma_f$. The equipartition theorem gives a measure of how much thermal energy is in each mode of a mechanical oscillator. The mean square vibration amplitude associated with a mode of oscillation at temperature $T$ is

$$\frac{1}{2}k_B T = \frac{1}{2}k\langle a^2\rangle \tag{S41}$$

where $k_B$ is Boltzmann's constant and $k$ is the spring constant of the mechanical oscillator. $\langle a\rangle^2$ is given by the integral over all frequencies of the single-sided pressure noise power spectral density $S_{PP}^{th}$ multiplied by the square of the mechanical response function

$$\langle a^2\rangle = \int_0^\infty |\mathrm{X}(\Omega)|^2 S_{PP}^{th} d\Omega. \tag{S42}$$

Under the assumption that the pressure noise spectrum is white, the corresponding $\mathrm{p_a}$ and $p_{\mathrm{th}}$ are given by

$$p_{\mathrm{a}} = \sqrt{S_{PP}^{a}} = \frac{2}{\sigma A}\sqrt{m_{\mathrm{eff}} k_B T \Gamma_f} \tag{S43}$$

for the fluid-induced noise and

$$p_{\mathrm{th}} = \sqrt{S_{PP}^{th}} = \frac{2}{\sigma A}\sqrt{m_{\mathrm{eff}} k_B T \Gamma_{\mathrm{int}}} \tag{S44}$$

for the intrinsic mechanical noise where $A$ is the detector area, and $\sigma = \frac{|\int P(r)Q(r)dA|}{(\max|P|)(\max|Q|)A}$ is the overlap integral between the displacement field, $Q$, and the pressure field, $P$, taken over the surface of the OMC receiver facing the impinging acoustic beam. In our case, $\sigma$ is calculated to be approximately 0.2 for a gap of 10 µm between the OMC transmitter and receiver from our 3D FEM model. To obtain Eqn. (S43) and Eqn. (S44), we have used the response function of the mechanical oscillator $X(\Omega) = a(\Omega)/P(\Omega)$, which is found by transforming Eqn. (S2) to Fourier space:

$$-\Omega^2 a + i\Omega\Gamma a + \Omega_m^2 a = \frac{F_{\mathrm{appl}}(\Omega)}{m_{\mathrm{eff}}} \tag{S45}$$

where $F_{\mathrm{appl}} = P\sigma A$. The response function, $X(\Omega)$, is then

$$X(\Omega) = \frac{a(\Omega)}{P(\Omega)} = \frac{1}{\frac{m_{\mathrm{eff}}}{\sigma A}\left(\Omega_m^2 - \Omega^2 + i\Omega(\Gamma_{\mathrm{int}} + \Gamma_f)\right)}. \tag{S46}$$

In an OMC ultrasound receiver, back-action noise due to radiation pressure arises as fluctuating forces on the mechanical element. The radiation-pressure force noise, $S_{FF}^{\mathrm{BA}}$, is given by

$$S_{FF}^{\mathrm{BA}} = \frac{\hbar^2\omega_0^2}{L_{\mathrm{OM}}^2} n_{\mathrm{cav}}\gamma \left[\frac{1}{(\gamma/2)^2 + (\Delta + \Omega)^2} + \frac{1}{(\gamma/2)^2 + (\Delta - \Omega)^2}\right] \tag{S47}$$

where $n_{\mathrm{cav}} = \frac{\gamma_w}{\Delta^2+(\gamma/2)^2}\frac{P_{\mathrm{in}}}{\hbar\omega_0}$ is the number of photons in the cavity [68]. The NEP is then

$$\mathrm{p_{BA}} = \sqrt{S_{PP}^{\mathrm{BA}}} = \frac{\hbar\omega_0}{L_{\mathrm{OM}}\sigma A}\sqrt{n_{\mathrm{cav}}\gamma\left[\frac{1}{(\gamma/2)^2 + (\Delta + \Omega)^2} + \frac{1}{(\gamma/2)^2 + (\Delta - \Omega)^2}\right]}. \tag{S48}$$

For the case where $\Delta = -\Omega$ and $\Omega \gg \gamma/2$, Eqn. (S48) reduces to

$$\mathrm{p_{BA}} = \sqrt{S_{PP}^{\mathrm{BA}}} = \frac{2\hbar\omega_0}{L_{\mathrm{OM}}\sigma A}\sqrt{\frac{1}{\gamma}\frac{\gamma_w}{\Omega^2}\frac{P_{\mathrm{in}}}{\hbar\omega_0}}. \tag{S49}$$

To determine the contribution to the NEP from shot noise, detector noise, and RF amplifier noise, we also need to know the optomechanical transfer function, $X_{OM}$, given by $X_{OM} = P_\Omega/a_{Rx}$. Starting from Eqn. (S18), Supplementary Note 9 gives a derivation of the output power of the receiver, $P_\Omega$, as a function of $\beta$ for the case where $\Delta = -\Omega$ and $\Omega \gg \gamma/2$. Under these approximations, the expression for $P_\Omega$ reduces to

$$P_\Omega(\Delta = -\Omega) = \frac{2\beta\gamma_w^2}{\gamma\Omega}\mathrm{P_{in}}. \tag{S50}$$

The expression for $X_{OM}$ is then

$$X_{OM}(\Delta = -\Omega) = \frac{P_\Omega(\Delta = -\Omega)}{a_{Rx}} = \frac{2\omega_0}{L_{\mathrm{OM}}}\frac{1}{\gamma}\left(\frac{\gamma_w}{\Omega}\right)^2\mathrm{P_{in}}. \tag{S51}$$

The contribution to the NEP from shot noise, $p_{\mathrm{SN}}$, is due to fluctuations in the number of detected photons. The single-sided power spectral density for shot noise is

$$S_{pp}^{SN} = \frac{2\hbar\omega_0 P_{\mathrm{det}}}{\eta_{\mathrm{qe}}} \tag{S52}$$

where $\hbar$ is Planck's constant divided by 2π, $\eta_{\mathrm{qe}}$ is the quantum efficiency given by $\eta_{\mathrm{qe}} = \frac{\hbar\omega_0 R}{e}$ where $R$ is the photodiode responsivity, and $P_{\mathrm{det}}$ is the optical power at the detector. The contribution $p_{\mathrm{det}}$ arises from noise sources present in the detector, which is captured by a noise-equivalent power ($\mathrm{NEP_{det}}$). The PSD corresponding to the detector $\mathrm{NEP_{det}}$ is then

$$S_{pp}^{\mathrm{det}} = \mathrm{NEP_{det}^2}. \tag{S53}$$

Converting to a NEP we have

$$p_{\mathrm{SN}} = \sqrt{S_{PP}^{SN}} = \frac{\gamma\Omega^2 L_{\mathrm{OM}}}{2P_{\mathrm{in}}\gamma_w^2\omega_0}\frac{1}{|\mathrm{X}(\Omega)|}\sqrt{\frac{2\hbar\omega_0 P_{\mathrm{det}}}{\eta_{\mathrm{qe}}}} \tag{S54}$$

and the corresponding $p_{\mathrm{det}}$ is

$$p_{\mathrm{det}} = \sqrt{S_{PP}^{\mathrm{det}}} = \frac{\gamma\Omega^2 L_{\mathrm{OM}}}{2P_{\mathrm{in}}\gamma_w^2\omega_0}\frac{1}{|\mathrm{X}(\Omega)|}\mathrm{NEP_{opt}}. \tag{S55}$$

We must also consider the noise associated with the RF amplifier after the photoreceiver, which has a noise equivalent pressure given by

$$p_{\mathrm{amp}} = \sqrt{S_{PP}^{\mathrm{amp}}} = \frac{\gamma \Omega^2 L_{\mathrm{OM}}}{2 P_{\mathrm{in}} \gamma_w^2 \omega_0} \frac{1}{|\mathrm{X}(\Omega)|} \frac{1}{G_P} \sqrt{S_{VV}^{\mathrm{amp}}} \tag{S56}$$

where $G_P$ is the conversion gain for the photoreceiver and $S_{VV}^{\mathrm{amp}} = 4 R_s k_B T F_n$ is the total input-referred noise for the RF amplifier expressed as a noise voltage spectral density, which includes the source's Johnson noise and the amplifier's excess noise, where $R_s$ is the source impedance and $F_n$ is the noise factor of the amplifier.

While the OMC receiver could be used in a system with a microwave RF source, there are additional noise sources that would need to be considered if an OMC transmitter is used to fully predict the behavior of the entire system. For example, temperature fluctuations will modify the mechanical frequency at which the OMC ultrasound transmitter oscillates, causing a fluctuating source frequency. We leave characterization of the noise properties of an OMC source to future work. Temperature fluctuations will also modify the mechanical resonance frequency for the OMC receiver, causing the optical power incident on the detector to fluctuate. This is another source of noise in the receiver, but given the broad linewidth for the mechanical cavity, it is expected for this source of noise to be negligible in our system. These temperature fluctuations will also modify the optical cavity resonance frequency in the OMC receiver, giving rise to thermo-refractive noise and thermal-mechanical noise, which we do include in our analysis of the NEP, as shown in Eqn. (S40).

**Supplementary Note 8. Approximating NEP from thermal shift**

In our OMC ultrasound receiver, temperature fluctuations will modify the optical cavity resonance frequency, indistinguishable from frequency shifts caused by mechanical motion, leading to additional thermal noise. The change in optical frequency is primarily due to the thermo-optic effect, where the refractive index of the silicon changes with temperature, and thermal expansion, where the physical dimensions of the cavity change with temperature. In Eqn. (S40), the corresponding NEP contributions are denoted by $\mathrm{p_{TR}}$ and $\mathrm{p_{TM}}$, respectively. Here we can consider that they give rise to a combined NEP that we denote as $p_{\mathrm{T}}$. Considering both of these effects, the optical resonance frequency shift, $\Delta\omega_0$, due to a temperature shift, $\Delta T$, is given by

$$\Delta\omega_0 = -\omega_0 \left( \frac{1}{n} \frac{dn}{dT} + \alpha_L \right) \Delta T \tag{S57}$$

where $n$ is the effective refractive index of the optical mode in the silicon cavity, $dn/dT$ is the thermo-optic coefficient of silicon, and $\alpha_L$ is the linear coefficient of thermal expansion of silicon. The thermo-optic coefficient for silicon is $1.94 \times 10^{-4}\ \mathrm{K}^{-1}$ while the linear coefficient of thermal expansion is $\sim 3 \times 10^{-6}\ \mathrm{K}^{-1}$ such that the thermo-optic effect dominates the frequency shift [69]. If we assume that the temperature fluctuations, $\Delta T$, follow a power spectral density, $S_{tt}$, then we can define an optical frequency noise power spectral density, $S_{\omega\omega}$, according to

$$S_{\omega\omega} = \omega_0^2 \left( \frac{1}{n} \frac{dn}{dT} + \alpha_L \right)^2 S_{tt}. \tag{S58}$$

The NEP, $p_{\mathrm{T}}$, is then given by

$$p_{\mathrm{T}} = \sqrt{S_{PP}^{T}} = \frac{1}{|\mathrm{X}(\Omega)|} \frac{L_{\mathrm{OM}}}{\omega_0} \sqrt{S_{\omega\omega}}. \tag{S59}$$

The expression for temperature fluctuations in a volume, $V$, is given by $\langle \Delta T^2 \rangle = k_B T^2/C$ where $C$ is the heat capacity given by $C = \rho c_p V$ where $c_p$ is the specific heat capacity at constant pressure [70]. The corresponding temperature fluctuation power spectral density for an OMC immersed in an aqueous solution, assuming a uniform temperature fluctuation over the optical mode volume, is given by

$$S_{tt} = \frac{4k_B T^2}{G} \frac{1}{1 + (\Omega_m \tau_t)^2} \tag{S60}$$

where $G = C/\tau_t$ is the thermal conductance, $\tau_t$ is the thermal time constant, and $C = \rho c_p V_{\text{eff}}$ where $V_{\text{eff}}$ is the effective volume of the optical mode in the OMC [70,71]. In this case, the thermal time constant, $\tau_t$, for an OMC immersed in an aqueous solution is determined as detailed in Supplementary Note 6.

**Supplementary Note 9. Derivation of $P_\Omega(\Delta = -\Omega)$**

For the optomechanical crystal ultrasound detector we set $\Delta = -\Omega$ and assume that $\Omega \gg \gamma/2$. In addition, we assume that $\beta \ll 1$ and, therefore, $J_0(\beta) \approx 1$ and $J_{\pm 1}(\beta) \approx \pm {}^{\beta}/_{2}$. Starting from Eqn. (S18), we can express $T_\Omega$ as:

$$\begin{aligned} T_\Omega &= 2J_0(\beta)J_1(\beta)\gamma_w^2 \left[ \text{Re}\left\{ \left( \frac{-i}{\left(\frac{\gamma}{2} + i\Omega\right)\left(\frac{\gamma}{2} - 2i\Omega\right)} + \frac{i}{\left(\frac{\gamma}{2} - i\Omega\right)\left(\frac{\gamma}{2}\right)} \right) e^{-i\Omega t} \right\} \right]. \\ &= 2J_0(\beta)J_1(\beta)\gamma_w^2 \left[ \text{Re}\left\{ \frac{2^2 i}{\gamma^2} \left( \frac{-1}{\left(1 + \frac{i\Omega}{\gamma/2}\right)\left(1 - 2i\frac{\Omega}{\gamma/2}\right)} + \frac{1}{\left(1 - i\frac{\Omega}{\gamma/2}\right)} \right) e^{-i\Omega t} \right\} \right]. \end{aligned} \tag{S61}$$

Let $X = \frac{\Omega}{\gamma/2}$:

$$\begin{aligned} T_\Omega &= 2J_0(\beta)J_1(\beta)\gamma_w^2 \left[ \text{Re}\left\{ \frac{4i}{\gamma^2} \left( \frac{-1 + iX + (1 + iX)(1 - 2iX)}{(1 + iX)(1 - 2iX)(1 - iX)} \right) e^{-i\Omega t} \right\} \right] \\ &= 2J_0(\beta)J_1(\beta)\gamma_w^2 \left[ \text{Re}\left\{ \frac{4i}{\gamma^2} \left( \frac{2X^2}{(1 + X^2)(1 - 2iX)} \right) e^{-i\Omega t} \right\} \right] \\ &= 2J_0(\beta)J_1(\beta)\gamma_w^2 \left[ \text{Re}\left\{ \frac{4i}{\gamma^2} \left( \frac{2X^2}{(1 - 2iX + X^2 - 2iX^3)} \right) e^{-i\Omega t} \right\} \right]. \end{aligned} \tag{S62}$$

We can assume that $\Omega \gg \gamma/2$ meaning that $X \gg 1$

$$\begin{aligned} T_\Omega &= 2J_0(\beta)J_1(\beta)\gamma_w^2 \left[ \text{Re}\left\{ \frac{8i}{\gamma^2} \left( \frac{X^2}{(X^2 - 2iX^3)} \right) e^{-i\Omega t} \right\} \right] \\ &= 2J_0(\beta)J_1(\beta)\gamma_w^2 \left[ \text{Re}\left\{ \frac{8i}{\gamma^2} \left( \frac{X^2(X^2 + 2iX^3)}{(X^4 + 4X^6)} \right) e^{-i\Omega t} \right\} \right] \\ &= 2J_0(\beta)J_1(\beta)\gamma_w^2 \left[ \text{Re}\left\{ \frac{8i}{\gamma^2} \left( \frac{X^4 + 2iX^5}{4X^6} \right) e^{-i\Omega t} \right\} \right]. \end{aligned} \tag{S63}$$

where we have omitted the $X^4$ term because $X \gg 1$. Further simplification gives:

$$T_{\Omega} = \frac{4J_0(\beta)J_1(\beta)\gamma_w^2}{\gamma^2}\left[\mathrm{Re}\left\{(iX^{-2} - 2X^{-1})e^{-i\Omega t}\right\}\right]$$
$$= \frac{4J_0(\beta)J_1(\beta)\gamma_w^2}{\gamma^2}[\mathrm{Re}\{(iX^{-2} - 2X^{-1})(\cos(\Omega t) - i\sin(\Omega t)\}]$$
$$= \frac{4J_0(\beta)J_1(\beta)\gamma_w^2}{\gamma^2}[-2X^{-1}\cos(\Omega t) + X^{-2}\sin(\Omega t)]. \quad \text{(S64)}$$

The transmitted amplitude oscillating at $\Omega$ is then

$$T_{\Omega} = A_c\cos(\Omega t) + A_s\sin(\Omega t) \quad \text{(S65)}$$

where

$$A_c = \frac{-8\gamma_w^2 J_0 J_1}{\gamma^2 X} = \frac{-4\gamma_w^2 J_0 J_1}{\gamma\Omega} \quad \text{(S66)}$$

and

$$A_s = \frac{4\gamma_w^2 J_0 J_1}{\gamma^2 X^2} = \frac{\gamma_w^2 J_0 J_1}{\Omega^2} \quad \text{(S67)}$$

meaning that

$$A_c = \frac{-4\Omega}{\gamma} A_s. \quad \text{(S68)}$$

The output power at $\Omega$ is then

$$P_{\mathrm{out},\Omega} = P_{\mathrm{in}}\sqrt{A_c^2 + A_s^2} \cong P_{\mathrm{in}}|A_c| = \frac{4\gamma_w^2 J_0 J_1}{\gamma\Omega}P_{\mathrm{in}} = 2\beta\frac{\gamma_w^2}{\gamma\Omega}P_{\mathrm{in}}. \quad \text{(S69)}$$

**Supplementary Note 10. Frequency scaling and optimization of thermal-acoustic noise**

To assess how the NEP scales with frequency, $f_m = \Omega_m/2\pi$, we need to consider the expression for the NEP associated with the thermal-acoustic noise of the medium, $p_a$. Through the fluctuation-dissipation theorem, this is given by $p_a = \frac{2}{\sigma A}\sqrt{m_{\mathrm{eff}} k_B T \Gamma_f}$ (see Supplementary Note 7), where $k_B$ is Boltzmann's constant, $T$ is the temperature, $\Gamma_f$ is the fluidic damping rate of the mechanical oscillator, $A$ is the detector area, and $\sigma = \frac{|\int P(r)Q(r)dA|}{(\max|P|)(\max|Q|)A}$ is the overlap integral between the displacement field, $Q$, and the pressure field, $P$, taken over the surface of the OMC receiver facing the impinging acoustic beam. Under the assumption of geometric scaling to achieve a higher operating frequency, where all dimensions are proportional to $\sim 1/f_m$, the detector area, $A$, scales as $\sim 1/f_m^2$ and the effective mass, $m_{\mathrm{eff}}$, scales as $\sim 1/f_m^3$. In the regime where the dominant fluidic loss channel is acoustic radiation loss, it can be assumed that $\Gamma_f$ scales as $\sim f_m$ [72-75]. Combining these scaling effects, the thermal-acoustic NEP, $p_a$, is expected to degrade approximately linearly with increasing operating mechanical frequency. This linear scaling is consistent with our ~150X higher operating frequency (4.6 GHz vs 30 MHz) corresponding to a ~150X increase in NEP ($0.2\ \mathrm{Pa}/\sqrt{\mathrm{Hz}}$ vs $1.3\ \mathrm{mPa}/\sqrt{\mathrm{Hz}}$).

While not applied in our analysis in this work, an OMC ultrasound platform does offer mechanisms to modify the thermal-acoustic noise floor, $p_{\mathrm{a}}$, itself. This could be accomplished through modifying the OMC ultrasound transmitter and detector pair design to improve the pressure field and displacement field overlap factor quantified by $\sigma$. If $\sigma$ increases to 1 for our approach presented here, our NEP reduces to $0.05\ \mathrm{Pa}/\sqrt{\mathrm{Hz}}$. A co-design between acoustic beamforming and mechanical modal displacement field profiles could be a viable path to achieve significantly lower thermal-acoustic NEP at ultra-high frequencies. Higher-order mechanical modes could also be utilized to achieve multi-GHz operation with a larger device footprint, which could potentially reduce the thermal-acoustic NEP by increasing $A$ and $\sigma$.

**Supplementary Note 11. Acoustic beam occlusion study**

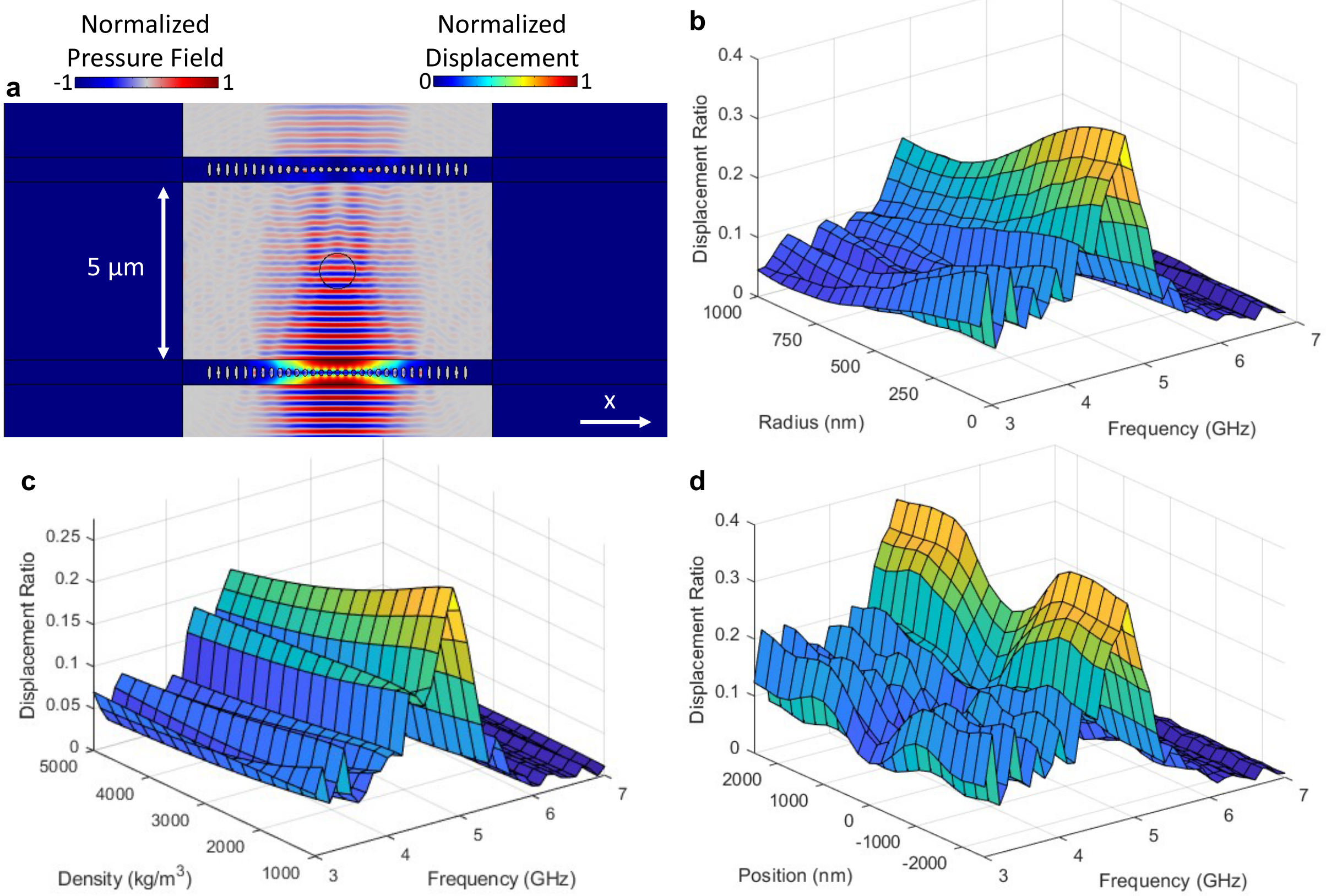

**Supplementary Figure 5: Object detection via acoustic beam occlusion.** (a) Snapshot of the 2D model to analyze the detection of a circular object placed between the OMC transducers. Plots of the OMC receiver and transmitter displacement ratio as a function of driving frequency and object (b) radius (c) density and (d) position relative to the center at $x = 0$.

We have carried out an initial assessment of the capabilities of the optomechanical ultrasound approach by using occlusion of the acoustic beam with an object as a proxy for imaging. As shown in Supplementary Fig. 5(a), a 2D FEM model was built with a circular object in-between the OMC transducers. The circular object has a nominal radius of 500 nm, a nominal density of 1050 kg/m$^3$, and a nominal position of $x = 0$, corresponding to the center of the OMCs. The speed of sound in the object is set to 1550 m/s and the acoustic attenuation is set to 4.34 dB/µm.

We monitor the displacement ratio, which is the ratio of the OMC receiver displacement to the OMC transmitter displacement. Plots of the displacement ratio as a function of operating frequency and object radius, density, and position are shown in Supplementary Fig. 5(b–d), respectively. As expected, the displacement ratio decreases with an increasing radius of the object and a decreasing position of the object relative to the center of the OMC pair. An increase in the object density also decreases the displacement ratio through the increased acoustic impedance.

## Supplementary Note 12. Dry cavity approach

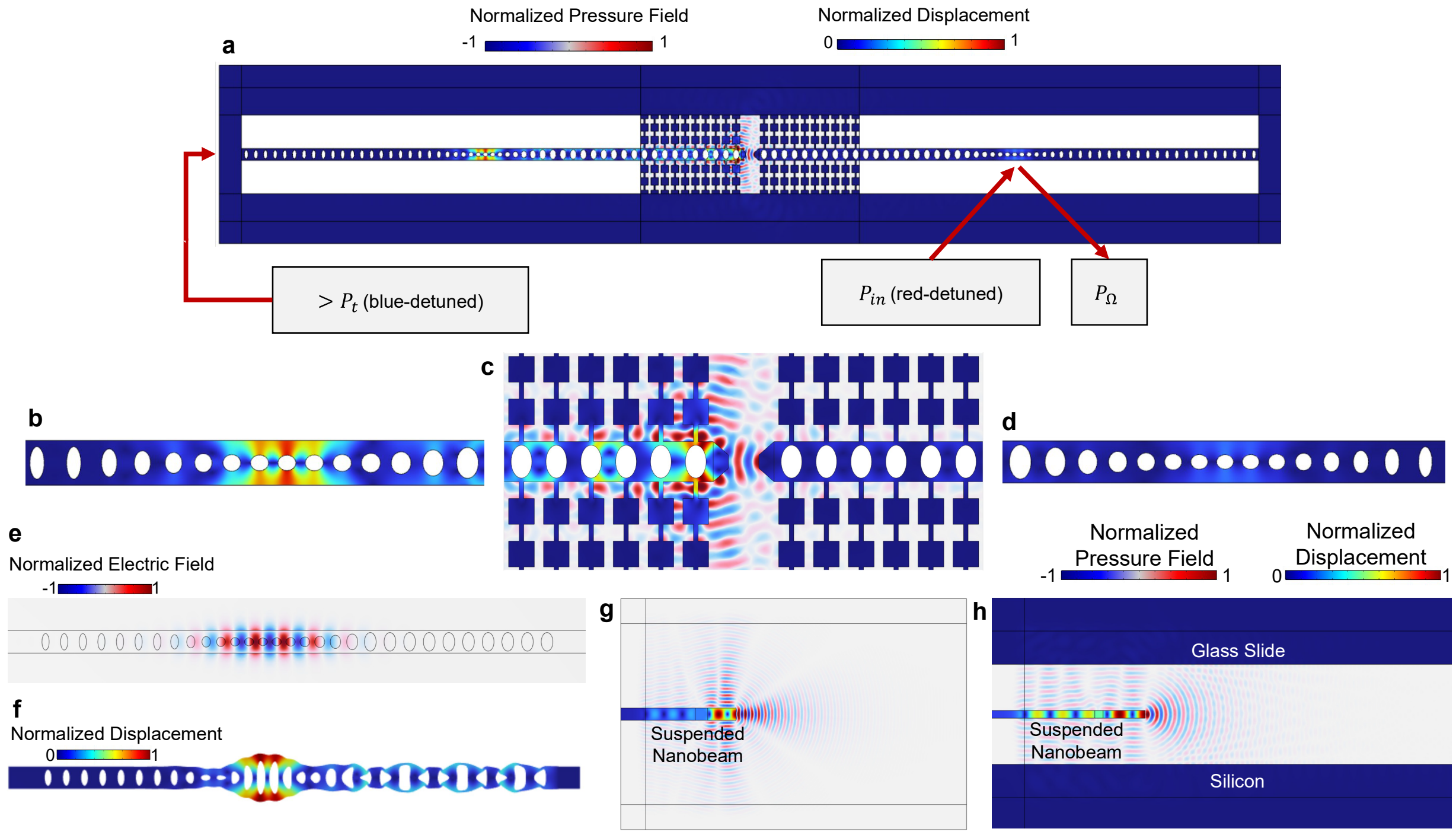

**Supplementary Fig. 6: Dry cavity approach.** (a) An overview of the dry cavity transduction scheme with an associated 2D FEM model. (b) The OMC ultrasound transmitter, in air, is driven into the regime of self-sustained optomechanical oscillation. The generated phonons leak into a phononic waveguide which then (c) routes the phonons to an aqueous solution. This generates a pressure field in the fluid, which ultimately induces mechanical displacement in (d) an OMC ultrasound receiver operating in air. The normalized electric and displacement fields for the OMC from a 3D FEM model is shown in (e) and (f), respectively. This approach works based on the generation of a pressure field in an aqueous solution from a driven suspended silicon nanobeam, as illustrated in the 2D FEM models shown in (g) (top-down view) and (h) (cross-sectional view).

As shown in Supplementary Fig. 6(a), here we present an additional design to the nano-optomechanical ultrasound platform where an analogous transduction process is utilized, but the OMC devices are separated from the aqueous solution such that they can operate in air. Phononic waveguiding is utilized to both route the generated phonons in the transmitter OMC (Supplementary Fig. 6(b)) to the aqueous solution (Supplementary Fig. 6(c)) producing the acoustic beam and to route the resulting displacement to the receiver OMC (Supplementary Fig. 6 (d)) for detection. This approach enables operation at an optical wavelength of 1550 nm without deleterious optical absorption in the fluid and therefore is able to rely on OMC designs that have already been experimentally characterized in the literature [76,77] with optical and mechanical modes as shown in Supplementary Fig. 6(e) and Supplementary Fig. 6(f), respectively. As can be seen, by design the ~5 GHz mechanical mode extends into the single mode phononic waveguide, while the optical mode has a bandgap at the operating frequency in the waveguide and thus is strongly confined to the cavity region. This approach has been demonstrated in several different optomechanical systems [76-79]. The phononic waveguide is tethered for mechanical support with a phononic crystal with a bandgap that prevents leakage into the substrate. The approach operates based on the ability to generate a pressure field in an aqueous solution from a driven suspended silicon nanobeam, as shown in Supplementary Fig. 6(g) and Supplementary Fig. 6(h).

The design where the OMC ultrasound transducers are directly submerged in the aqueous solution, thereby emitting and absorbing phonons directly into and from the fluid, does have specific advantages, such as an increased ability to engineer the acoustic beam properties through tailoring the mechanical mode sidewall profile and harnessing the enhanced thermal properties of the fluid. However, in the dry cavity approach, given the separation of the OMC device from the aqueous solution, we have the potential to achieve a higher optical quality factor, a higher mechanical quality factor, and a lower effective mass. As expected, the dry cavity approach is limited by diffraction losses, which can potentially be modified through additional design of the phononic waveguide and its taper to engineer an optimal transition region. For example, there is potential for an acoustic inverse taper to expand the mode field diameter of the acoustic wave in the water, thereby achieving a larger Rayleigh range. It is also important to note that there may be applications in ultrasound imaging and sensing that benefit from a diverging beam; the most suitable application may, in part, then be driven by the ease in which a diffracting, collimated, or focused acoustic beam can be experimentally realized.